\documentclass[fleqn,usenatbib]{mnras}

\usepackage{float}
\usepackage[T1]{fontenc}
\usepackage[utf8]{inputenc}
\usepackage{float}
\usepackage{url}
\usepackage{multirow}
\usepackage{ae,aecompl}
\usepackage[dvipsnames]{xcolor}

\usepackage{graphicx}	% Including figure files
\usepackage{amsmath}	% Advanced maths commands
\usepackage{amssymb}	% Extra maths symbols

\usepackage{color}

\title[Magnetic field generation from primordial black hole distributions]{Magnetic field generation from primordial black hole distributions}
\author[Araya et. al.]{I. J. Araya,${}^{1}$\thanks{E-mail: araya.quezada.ignacio@gmail.com} M. E. Rubio${}^{2,3}$, M. San Martín${}^{4}$, F. A. Stasyszyn${}^{2,3}$, N. D. Padilla${}^{4,5}$,
\newauthor{J. Magaña${}^{4,5}$ and J. Sureda${}^{4,5}$}
\\
${}^{1}$Instituto de Ciencias Exactas y Naturales (ICEN), Facultad de Ciencias, Universidad Arturo Prat,\\
Avenida Arturo Prat Chac\'on 2120, 1110939, Iquique, Chile\\
${}^{2}$Instituto de Astronomía Teórica y Experimental (IATE-CONICET), Laprida 854, Córdoba, Argentina\\
${}^{3}$Observatorio Astronómico de Córdoba (OAC-UNC), Laprida 854, Córdoba, Argentina\\
${}^{4}$Instituto de Astrofísica, Pontificia Universidad Católica de Chile, Avda. Vicuña Mackenna 4860, Santiago, Chile\\
${}^{5}$Centro de Astro-Ingeniería, Pontificia Universidad Católica de Chile, Vicuña Mackenna 4860, Santiago, Chile
}

\date{Accepted XXX. Received YYY; in original form ZZZ}

\pubyear{2020}

\begin{document}
\label{firstpage}
\pagerange{\pageref{firstpage}--\pageref{lastpage}}
\maketitle

% Abstract of the paper
\begin{abstract}
We introduce a statistical method for estimating magnetic field fluctuations generated from primordial black hole (PBH) populations. To that end, we consider monochromatic and extended Press-Schechter PBH mass functions, such that each constituent is capable of producing its own magnetic field due to some given physical mechanism. Assuming linear correlation between magnetic field fluctuations and matter over-densities, our estimates depend on the mass function, the physical field generation mechanism by each PBH constituent, and the characteristic PBH separation. After computing the power spectrum of magnetic field fluctuations, we apply our formalism to study the plausibility that two particular field generation mechanisms could have given rise to the expected seed fields according to current observational constraints. The first mechanism is the Biermann battery and the second one is due to the accretion of magnetic monopoles at PBH formation, constituting magnetic PBHs. Our results show that, for monochromatic distributions, it does not seem to be possible to generate sufficiently intense seed fields in any of the two field generation mechanisms. For extended distributions, it is also not possible to generate the required seed field by only assuming a Biermann battery mechanism. In fact, we report an average seed field by this mechanism of about $10^{-47}$ G, at $z = 20$. For the case of magnetic monopoles we instead assume that the seed values from the literature are achieved and calculate the necessary number density of monopoles. In this case we obtain values that are below the upper limits from current constraints. 
\end{abstract}

% Select between one and six entries from the list of approved keywords.
% Don't make up new ones.

%%%%%%%%%%%%%%%%%%%%%%%%%%%%%%%%%%%%%%%%%%%%%%%%%%

%%%%%%%%%%%%%%%%% BODY OF PAPER %%%%%%%%%%%%%%%%%%

\section{Introduction}

There is no doubt that magnetic fields are ubiquitous in any astrophysical system, constituting one of the main characters for the description of the structure, dynamics and evolution of the Universe at all its scales. Magnetic fields are present in both large and small scale structures \citep{Battaner97,Pandey15}, as well as in the description of how particle acceleration operates throughout the IGM \citep{Kronberg02}. They also play an unavoidable role in the evolution of primordial plasma in the Early Universe and in the propagation of cosmic rays in clusters of galaxies, being a relevant and subtle influence in galaxy formation processes. The study and observation of magnetic fields at high redshift is, nevertheless, still challenging, as existing fields before the CMB formation might leave imprints from their impact on plasma physics that could hopefully be observable in the upcoming years. 

Primordial magnetic field astronomy represents one of the most intense and challenging research areas in Cosmology, both from the theoretical point of view, as well as from observations and numerical simulations \citep{Widrow02,Widrow11,Beck2013}. It has opened an exciting window into the Early Universe, particularly in the period between inflation and recombination. Also, primordial magnetic fields have a relevant impact on CMB anisotropies \citep{Barrow97, Subramanian06}, being possible to constrain magnetic field strength from Big Bang nucleosynthesis \citep{Grasso96}. However, its origin, dynamical evolution and its amplification/re-generation mechanisms are still not clear, which motivated a broad set of intriguing hypotheses trying to understand them \citep{Kandus11, Subramanian16, Yamasaki12, Barrow07}. Many of the suggested magnetic field generating mechanisms require a previous ``seed field'', which gives rise to many open questions, like (i) How were magnetic seed fields created?; (ii) Which mechanisms govern the amplification from seed fields to more intense fields over time?; (iii) How is the large-scale magnetic field distribution organized and why? Different approaches to these fundamental questions have helped to provide some answers that are consistent with observations. Following the basic first principles of Magnetohydrodynamics, it is clear that if no magnetic field is initially present, no magnetic field could ever have been evolved \citep{Moffatt78}. It is expected that magnetic seed fields could be created from the Big Bang (although still it is not clear under which physical mechanism), but it is also believed that they could have been created in the Early Universe, before the first generation of galaxies. In such case, it seems to be probable that battery-like processes could have taken place, under certain particular conditions. 

Other generation mechanisms could have taken place during the reheating phase, the QCD phase transitions, the electroweak phase transitions, or prior to recombination. These speculations motivated other intriguing non-standard models suggesting a sort of classification of different analytical models in two main groups: the first one is for models with an ``astrophysical'' nature, while the second one has a more ``cosmological'' viewpoint. The first category includes Biermann-battery-like processes generated from intergalactic shocks, supernova implosions, stellar winds and galactic jets of magnetized plasma, among others \citep{Grasso01,Widrow02,Widrow11,Dar05}. Instead, cosmological models involve Early universe phase transitions, magnetic helicities, hypercharge and hypermagnetic field generation and cosmological second-order vector perturbations \citep{Quashnock89,Banerjee03,Joyce97,Giovannini98,Durrer06,Saga15}.

Lower bounds estimated from observations on extragalactic scales suggest that the existing fields could be remnants from primordial fields originated cosmologically rather than astrophysically. Following the analytical models that support these hypotheses, it is found that magnetic seed fields seem to be too weak in strength when compared to observations at different scales, clearly suggesting that some amplification mechanism needs to be present \citep{Beck96,Brandemburg07,Brandenburg18}. 

Although it has not yet been possible to model the mechanism for creating cosmological magnetic fields without resorting to ``beyond-standard-model'' ingredients \citep{Blas11}, they have been detected at several different scales, ranging from planetary scales to that of galactic clusters \citep{Turner88,Jasche15,Widrow11,Brandenburg050}. The strongest observational evidence for the existence of cosmological magnetic fields comes from the Zeeman effect, the synchotron radiation and the Faraday rotation \citep{Widrow02}. For strong fields, the Zeeman effect suffices as a probe, but not so for fields at the galactic scale. Instead, synchotron radiation (emitted from electrons moving around intense magnetic field lines) turns out to be useful for galactic observations. Finally, a Faraday rotation signal can be read from its frequency dependence: fields on supercluster scales are actually difficult to estimate, ranging from nG to $\mu$G orders of magnitude \citep{Widrow11}.

Observations of large-scale magnetic fields offer clear insights about regular ordered patterns of the field lines, suggesting that mean-field dynamo processes are responsible for their order and structure, as well as the existence of additional transport processes carrying magnetic energy into huge regions of space \citep{Govoni17,Han17,Arlen12,Bonafede10,Clarke01}. Magnetic fields cannot be directly observed, so their impact on radiation processes needs to be considered (see, for instance, \cite{Rybicki79} and references therein).
In addition, observations of magnetic fields in voids provide bounds on their strength, depending on the analytical model: from the simplest ones, it is possible to obtain lower bounds, although when improving such models a bounded range of magnetic field strength values can be provided, ranging from $10^{-25}$ to $10^{-15}$ nG \citep{Takahashi13,Essey11,Tavecchio10,Neronov10}. Other authors argue magnetic field strengths between $10^{-16}$ G and $10^{-15}$ G \citep{Hubble29,Hubble31,Einstein16,Einstein15,Bull16}. Magnetic fields from astrophysical voids are relevant since they could evidence truly cosmological magnetic fields, which could have served as seeds for magnetic fields in lower scales (such as galactic fields). Observations in galaxy clusters yield values of order $10^{-6}$ G \citep{Boulanger18}. Interest in primordial magnetic fields generated during inflation has recently increased. This scenario has driven the search for  phenomenologically viable mechanisms to explain the observed magnetic fields in a broad set of scales \citep{Demozzi09,Brandenburg050,Grasso01,Ferreira13,Green16}. The latest data on reionization and the observed UV luminosity function of high-redshift galaxies place limits on the magnetic field strength due to its impact on the reionization process.

Surprisingly, it has been pointed out by \citet{Saga2020} that magnetic fields can act as a non-Gaussian source for the cosmic density fluctuation field, which can impact the formation of primordial black holes (PBHs, e.g. \cite{Khlopov_2010}), leading to an enhancement with respect to the Gaussian fluctuation case. Conversely, one could wonder whether or not PBHs could generate seed magnetic fields which could explain current measurements in cosmological scales.

In this work, we study the generation of a cosmic magnetic field by PBHs, which are assumed to comprise a sizeable fraction of the dark matter in the universe \citep{Belotsky_2014}, through two different mechanisms. The first one is commonly referred to as the Biermann battery mechanism \citep{Biermann:1950}, through which a magnetic field is generated due to the presence of an accretion disk around the black hole, under certain thermodynamical conditions on pressure and gas density gradients, other than geometrical assumptions. The second mechanism considers the possibility that PBHs are magnetic black holes which have a net magnetic charge as a
result of the abundance of magnetic monopoles in collapsing regions during PBH formation \citep{Maldacena20}. In order to get estimates for the net magnetic field produced in both models, we assume that each PBH generates a magnetic field whose magnitude, measured at a distance $r$ from the center, depends mainly on the PBH mass (in the particular case of the Biermann battery mechanism, this implies assuming that PBHs are \textit{near-extremal}, which is well-motivated and discussed
in \cite{Pacheco20,DeLuca--2020}). Once we have the magnitude of the magnetic field for an individual PBH due to the corresponding process, we consider the PBH spatial clustering statistics, encoded in the dark matter over-density field and in the distribution of PBH masses for which we adopt both monochromatic and PS distributions, in order to find the power spectrum of the resulting cosmic magnetic field, and thus its average energy density. By assuming the highest possible number density of PBHs (i.e., that PBHs comprise all of the dark matter in the universe) and a maximal spin parameter for each PBH in the case of the Biermann battery, our results constitute upper bounds for the average magnetic field produced by this mechanism from PBH populations. For the case of magnetic PBHs, we constrain the number density of magnetic monopoles required to source the magnetic field up to the estimated values and, therefore, the fraction in magnetic monopoles of the cosmic matter density.

This paper is organized as follows. In section \ref{sec:2}, we review the required preliminary knowledge about the matter distribution and its clustering statistics, including the matter power spectrum, the variance of the matter field at a certain scale, and the PBH mass functions considered in this work. In section \ref{sec:3}, we discuss two different magnetic field generation mechanisms, which can endow the PBHs with a magnetic field, namely; the Biermann battery mechanism and the absorption of magnetic monopoles by PBHs during their formation, turning them into magnetic black holes. In section \ref{sec:4}, we present our general
prescription for computing the magnetic power spectrum and the cosmic magnetic
energy density, considering that the clustering of PBHs is described by the
matter power spectrum, and that each individual PBH generates a magnetic field by some (unspecified) physical process. In section \ref{sec:5}, we apply our formalism for the computation of magnetic power spectra and characteristic field strength to the two aforementioned mechanisms (namely, Biermann battery and magnetic black holes), obtaining the corresponding numerical results. Finally, general conclusions and a discussion on future research avenues are included in section \ref{sec:6}.

\section{Matter clustering statistics and PBH mass functions}
\label{sec:2}

In this section we review some preliminary knowledge on matter density field statistics and its fluctuations. We refer the reader to \cite{PeacockBook} for more details, specially the sections about the power spectrum and the correlation function of density fluctuations. We also include here a discussion on PBH mass functions and matter power spectra, particularizing on the ones that shall be considered along this work.

\subsection{The cosmic over-density field}

We start by considering the \textit{overdensity field}, $\delta(\vec{x})$, which is defined as the relative excess density at a given point $\vec{x}$ with respect to the average
(matter) cosmic density, $\overline{\rho}$; namely,
\begin{equation}
\delta(\vec{x}) =\frac{\rho(\vec{x}) -\overline{\rho}}{\overline{\rho}},
\end{equation}
being $\rho(\vec{x})$ the matter density field at position $\vec{x}$. Given that it is generally assumed that $\delta(\vec{x})$ has support over the whole cosmic volume (which we denote here by $V$), one is often interested in analyzing the behavior of $\delta(\vec{x})$ only over a particular size-scale (say a ball $\mathcal{B}_R$ of radius $R > 0$, which is completely defined by means of the  particular problem of interest). Thus, it is natural to define a \textit{smoothed overdensity field} as the convolution of $\delta(\vec{x})$ with some window scalar function $W_R(\vec{x})$ with compact support $\Omega\subseteq\mathcal{B}_R$ as follows:
\begin{equation}
\delta_{R}(\vec{x}) = \frac{1}{|\mathcal{B}_R|}
\int_{\mathbb{R}^3}{W_{R}\left(\vec{x}
-\vec{y}\right)\,\delta(\vec{y})\; \mbox{d}^{3}\vec{y}},
\end{equation}
where $|\mathcal{B}_R|$ is the size of $\mathcal{B}_R$. The function $W_{R}$ is chosen in order to average out the density field within the ball. For instance, $W_{R}$ could be a spherical top-hat function. From the above smoothed overdensity field, it is possible to define its \textit{variance}, $\sigma^2(R)$, over \textit{arbitrary} balls of radius $R$ as
\begin{equation}
\sigma^2(R) = 
\left\langle \left\vert \delta_{R}(\vec{x})\right\vert
^{2}\right\rangle_V
= \frac{1}{|V|}\int_{\mathbb{R}^3}{\left\vert \delta_{R}(\vec{x}) \right\vert^{2}\mbox{d}^{3}\vec{x}}\,,
\end{equation}
where $|V|$ is the size of $V$. The above function captures all the statistical behavior of the density fluctuations, averaged over the \textit{whole} cosmic space. It represents the \textit{relative variance} of the density field (with respect to the average) when smoothed over
spheres of radius $R$. 

It is automatically noticed that the larger the value of $R$, the smaller the variance, as
within larger scales the density field tends to be more homogeneous. Nonetheless, $\sigma^2(R)$ can be computed with relative ease, making use of suitable properties of the Fourier transformation. In effect, denoting by $\widehat{f}(\vec{k})$ the Fourier
transform of $f(\vec{x})$ one gets, by the convolution formula,
\begin{align}
\widehat{\delta}_R(\vec{k}) & = \frac{1}{|\mathcal{B}_R|}\left[\widehat{\int_{\mathbb{R}^3}{W_{R}\left(\vec{x}-\vec{y}\right)\,\delta(\vec{y})\; \mbox{d}^{3}\vec{y}}}\right](\vec{k}) \nonumber \\
&  = \widehat{W}_R (\vec{k})\; \widehat{\delta}(\vec
{k}),
\end{align}
where $\widehat{W}_R(\vec{k})$ is the Fourier transform of $W_R(\vec{x})$ divided by $|\mathcal{B}_R|$ (in order for it to be a dimensionless window function). Now, using Parseval's theorem, we have
\begin{align}
\sigma^{2}(R)  & =  \frac{1}{|V|}\int_{\mathbb{R}^3}{\left\vert \delta_{R}(\vec{x})\right\vert^{2}\mbox{d}^{3}\vec{x}}\nonumber\\
&  = \frac{1}{|V|}\int_{\mathbb{R}^3}{\left\vert\widehat{\delta}_R(\vec{k})\right\vert^{2} \mbox{d}^{3}\vec{k}}\nonumber \\
&  = \frac{1}{|V|}\int_{\mathbb{R}^3}{\left\vert\widehat{W}_R(\vec{k}) \right\vert^{2}\,\left\vert \widehat{\delta}(\vec{k})  \right\vert^{2}\,\mbox{d}^{3}\vec{k}}.
\end{align}
Thus, defining the \textit{power spectrum} of density fluctuations as
\begin{equation}
P(\vec{k}) = \frac{1}{|V|} \left\vert \widehat{\delta}(\vec{k}) \right\vert^{2},
\end{equation}
we can express
\begin{equation}
\sigma^{2}(R)  = \int_{\mathbb{R}^3}{\left\vert \widehat{W}_R(\vec{k}) \right\vert^{2}P(\vec{k})\,\mbox{d}^{3}\vec{k}}.
\end{equation}
Finally, assuming \textit{cosmic isotropy} together with a real window function in Fourier space, we get
\begin{equation}
\sigma^{2}(R,z)  =4\pi \int_{0}^{\infty}{\widehat{W}_R^{2}(
\vec{k}) P_{\mbox{\tiny{CDM}}}(\vec{k},z)\,  k^{2}\,\mbox{d} k},
\end{equation}
where $P_{\mbox{\tiny{CDM}}}(k,z)$ is the matter power spectrum at redshift $z$, coming from the concordance $\Lambda$CDM model.

\subsection{PBH mass functions}

Several mechanisms have been proposed for the formation of PBHs. Commonly, PBHs can be formed from the collapse of density perturbations in the Early Universe, during the radiation domination epoch \citep{Zeldovich1966, Hawking1971, Carr_Hawking:1974}. The standard picture assumes that a PBH  forms when a fluctuation exceeding a threshold critical density re-enters the horizon. Under this scenario, the resulting PBH mass should be proportional to the horizon mass in such epoch ($M_{\tiny{\text{PBH}}}\sim M_{H}$). 

It is useful to introduce the \textit{mass function} $dn/dM$ as the \textit{comoving number density of PBHs per unit mass}. In other words, the (comoving) number density (i.e., number of objects per (comoving) cubic length) of PBHs with masses within a given range of interest $M\in [M_{i},\,M_{f}]$, is determined by integrating the mass function over such whole range; namely,
\begin{equation}
n = \int_{M_{{\tiny{i}}}}^{M_{{\tiny {f}}}}{\frac{\mbox{d}n}{\mbox{d}M}\,\mbox{d}M}.
\end{equation}

Some models commonly assume that PBH mass functions are \textit{monochromatic}; i.e., all the PBHs have the same mass \citep{Niemeyer:1998PhRvL,Jedamzik:1999PhRvD}. Nevertheless, other formation scenarios lead to \textit{extended} PBH mass distributions such as power laws or lognormal forms, among others \citep{Carr:2018,Byrnes:2019,Carr:2020a,Carr:2020b,DeLuca:2020}. Here we shall explore the consequences of a particular type of extended mass functions, alongside monochromatic ones, as explained in what follows.

Very recently, new extended PBH mass functions have been proposed, employing a modified Press-Schechter approach for the collapse of energy density fluctuations with a Gaussian probability
distribution \citep{Sureda:2020}. In that work, the authors consider a primordial power spectrum with a broken power law, whose spectral index is red for wave-numbers less than a certain pivot scale $k_{\tiny{\mbox{piv}}}$ (i.e, larger scales) and blue for wave-numbers greater than $k_{\tiny{\mbox{piv}}}$ (smaller scales). 
\begin{figure}
    \centering
    \includegraphics[scale=0.45]{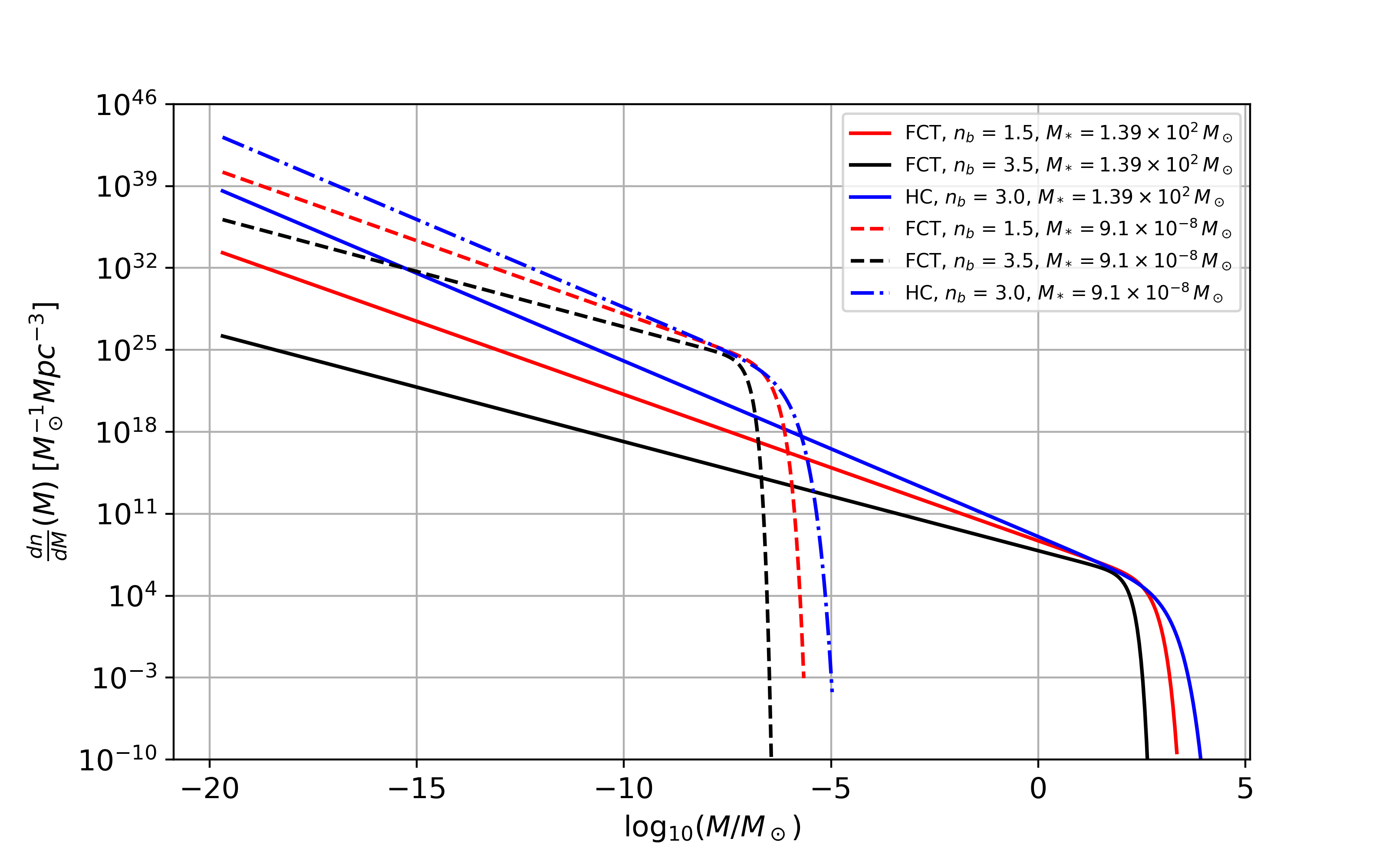}
    \caption{PBH extended mass functions for the different scenarios considered in this work. Solid lines correspond to mass functions with characteristic mass $M_* = 1.39\times 10^{2}\, M_\odot$, while dashed/dot-dashed lines correspond to $M_* = 9.1\times 10^{-8}\, M_\odot$. Blue curves correspond to the horizon crossing (HC) scenario, with spectral blue index $n_b = 3.0$. Red curves correspond to the fix conformal time (FCT) scenario, with $n_b = 1.5$ and black curves correspond also to a FCT formation scenario, with $n_b = 3.5$.}
    \label{fig:Extended-Mass-Functions}
\end{figure}
Two different scenarios for PBH formation are considered in \cite{Sureda:2020} in order to construct the mass functions: (i) a \textit{fixed conformal time} scenario (FCT) and (ii) a \textit{horizon crossing} (HC) scenario. In (i), PBHs of different masses are formed at the same time (or, more precisely, during a narrow time interval compared with the age of the universe). A possible mechanism of creation could have been for instance through ``bubble-like'' collisions in the Early Universe, under certain conditions which depend on the background solution, fixed by the cosmic dynamics \citep{Hawking82,Gleiser98,Ferrer19,Garriga16}. Instead, for the HC scenario, PBHs are formed at different times, when the size of the fluctuation matches the Hubble radius; i.e., when such scale crosses the horizon \citep{Green97,Green99,Green04,Green16}. Therefore, this mechanism restricts PBH formation to those scales of fluctuations which enter the horizon at a particular time. 

The aforementioned extended mass functions obtained through Press-Schechter theory are basically determined by two different parameters: a \textit{characteristic mass} $M_{*}$ (the one where the mass function starts an exponential \textit{cut-off}) and the blue index $n_{b}$. Using as constraint the mass function of supermassive black holes and some observational bounds at different mass ranges coming from monochromatic PBH mass functions, \cite{Sureda:2020} found allowed regions for the mass function parameters where all dark matter can be composed by PBHs. Throughout this work, we considered as fiducial characteristic masses the values $M_* = 1.39\times 10^2\,M_{\odot}$ and $M_* = 9.1\times 10^{-8}\,M_{\odot}$ for both scenarios. Also, for the FCT mechanism we considered two different cases: $n_b = 1.5$ and $n_b = 3.5$, while for the HC scenario we fixed the blue index to $n_b = 3.0$. Figure \ref{fig:Extended-Mass-Functions} shows the mass dependence on the comoving number density of PBHs in both scenarios under the chosen parameter constraints, where an abundance of low-mass PBHs is observed.

\subsection{Characteristic distance between equal-mass PBHs}
In order to understand the spatial distribution between equal-mass PBHs, and according to the extended mass functions introduced before, we find it useful to introduce the \textit{characteristic distance} between equal-mass PBHs as
\begin{equation}\label{eq:charact-distance}
d(M)  =\left(  \int_{M}^{\infty}{\frac{\mbox{d}n}{\mbox{d}M^{\prime}}\,\mbox{d}M^{\prime}}\right)^{-\frac{1}{3}}.
\end{equation}

Although this definition of $d(M)$ is \textit{non-local} in $M$, it has a direct interpretation. Firstly, it defines the comoving number density for all objects with masses which are greater than or equal to $M$ (given by the integral within the parenthesis). Secondly, it considers the comoving number density to compute the corresponding comoving average separation which, as usual, is given by $n^{-1/3}$, where $n$ is the comoving PBH number density. Also, $d(M)$ has the desired property that it is an increasing function of $M$, which means that, as heavier objects are less common in the universe, they have a greater (average) separation. 
\begin{figure}
    \centering
    \includegraphics[scale=0.45]{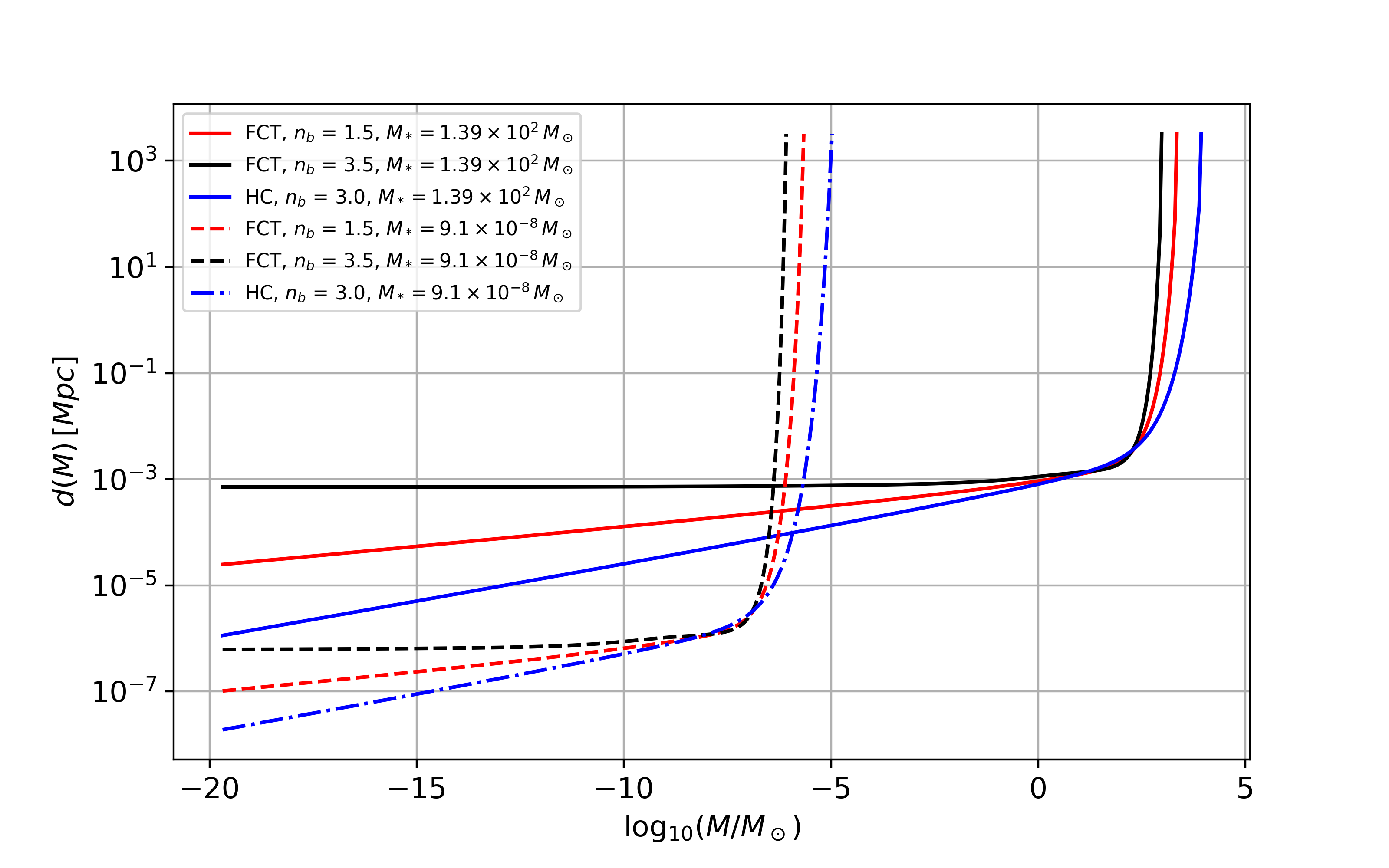}
    \caption{Characteristic distance function for the different scenarios considered in this work, assuming that PBHs constitute all the dark matter. Solid lines correspond to mass functions with characteristic mass $M_* = 1.39\times 10^{2}\, M_\odot$, while dashed/dot-dashed lines correspond to $M_* = 9.1\times 10^{-8}\, M_\odot$. Blue curves correspond to the horizon crossing (HC) scenario, with spectral blue index $n_b = 3.0$. Red curves correspond to the fix conformal time (FCT) scenario, with $n_b = 1.5$ and black curves correspond also to a FCT formation scenario, with $n_b = 3.5$.}
    \label{fig-charac-distance}
\end{figure}

\begin{table}
	\centering
	\begin{tabular}{ccc}
		\hline
		PBH Mass $[M_\odot]$ & $d_{\tiny{\mbox{mon}}}$ (FCT) [Mpc]  & $d_{\tiny{\mbox{mon}}}$ (HC) [Mpc]\\
		\hline\hline
		$3.5\times 10^{-17}$ & $2.04\times 10^{-35}$ & $4.27\times 10^{-21}$\\
		$10^{-15}$ & $6.23\times 10^{-35}$ & $1.3\times 10^{-20}$\\
		$10^{-13}$ & $2.89\times 10^{-34}$ & $6.06\times 10^{-20}$\\
		$5\times 10^{-11}$ & $2.29\times 10^{-33}$ & $4.81\times 10^{-19}$\\
		\hline
	\end{tabular}
	\caption{Characteristic distance for different PBH monochromatic distribution masses, considering the FCT and HC formation scenarios. For the calculation we set $f_{\tiny{\mbox{PBH}}}=1$ and the values for $\rho_{\tiny{\mbox{DM}}}$ for both scenarios were extracted from \citet{Sureda:2020}.}
	\label{tabla:dist-caract-monocromatic}
\end{table}

As shown in Figure \ref{fig-charac-distance}, as soon as PBH masses approach the minimum mass $M\rightarrow M_{i}$ (which corresponds to the evaporated mass, by Hawking radiation, at a particular redshift $z$), the characteristic distance tends to the \textit{average comoving separation} obtained considering \textit{all} the PBH population at that $z$. Although this may seem like a problem (as one should only be interested in those objects around
mass scale $M$), this is rather reasonable, since \textit{lower} mass objects are increasingly \textit{more abundant} (by several orders of magnitude). Then, for small PBH mass, the greatest contribution indeed comes from such mass scale.

Finally, we notice that in the case of a monochromatic mass function characterized only by one PBH mass, namely
\begin{equation}
    \left(\frac{dn}{dM}\right)_{\tiny{\mbox{mon}}} = n\,\delta(M - M_{\tiny{\mbox{PBH}}}),
\end{equation}
being $M_{\tiny{\mbox{PBH}}}$ the PBH mass, the corresponding characteristic distance simply reduces to
\begin{equation}
    d_{\tiny{\mbox{mon}}}(M) = n^{-1/3}.
\end{equation}
Here, $n$ is the comoving number density of PBHs, which can be expressed as
\begin{equation}
    n = \frac{f_{\tiny{\mbox{PBH}}}\,\rho_{\tiny{\mbox{DM}}}}{M_{\tiny{\mbox{PBH}}}},
\end{equation}
where $f_{\tiny{\mbox{PBH}}}$ is the fraction of dark matter energy density contribution coming from PBHs and $\rho_{\tiny{\mbox{DM}}}$ is the dark matter energy density. As an example, we compute the characteristic distance for monochromatic mass distributions, in a range of masses in which it is expected that all PBH dark matter has been formed (see Table \ref{tabla:dist-caract-monocromatic}). In particular, \citet{Dasgupta2020} estimate that the minimum mass of PBHs from which all dark matter may come (i.e., $f_{\tiny{\mbox{PBH}}}=1$) is $M_{\tiny{\mbox{min}}} \approx 10^{-16}\,M_\odot$, while the maximum mass is expected to be around $M_{\tiny{\mbox{max}}} \approx 5\times 10^{-11}\,M_\odot$ \citep{Smyth2020}. However, there are currently other constraints in dispute, such as NSs capturing or GRB femtolensing \citep{Carr:2020a}, which could make that range slightly vary. For instance, taking GRB femtolensing effects, the minimum mass would shift to $10^{-15}\, M_\odot$, and up to $10^{-14}\, M_\odot$. Also, \citet{Montero19} estimate that all dark matter could be in PBHs with masses from $3.5 \times 10^{-17}\,M_\odot$ to $4 \times 10^{-12}\,M_\odot$.

\begin{figure}
    \centering
    \includegraphics[scale=0.45]{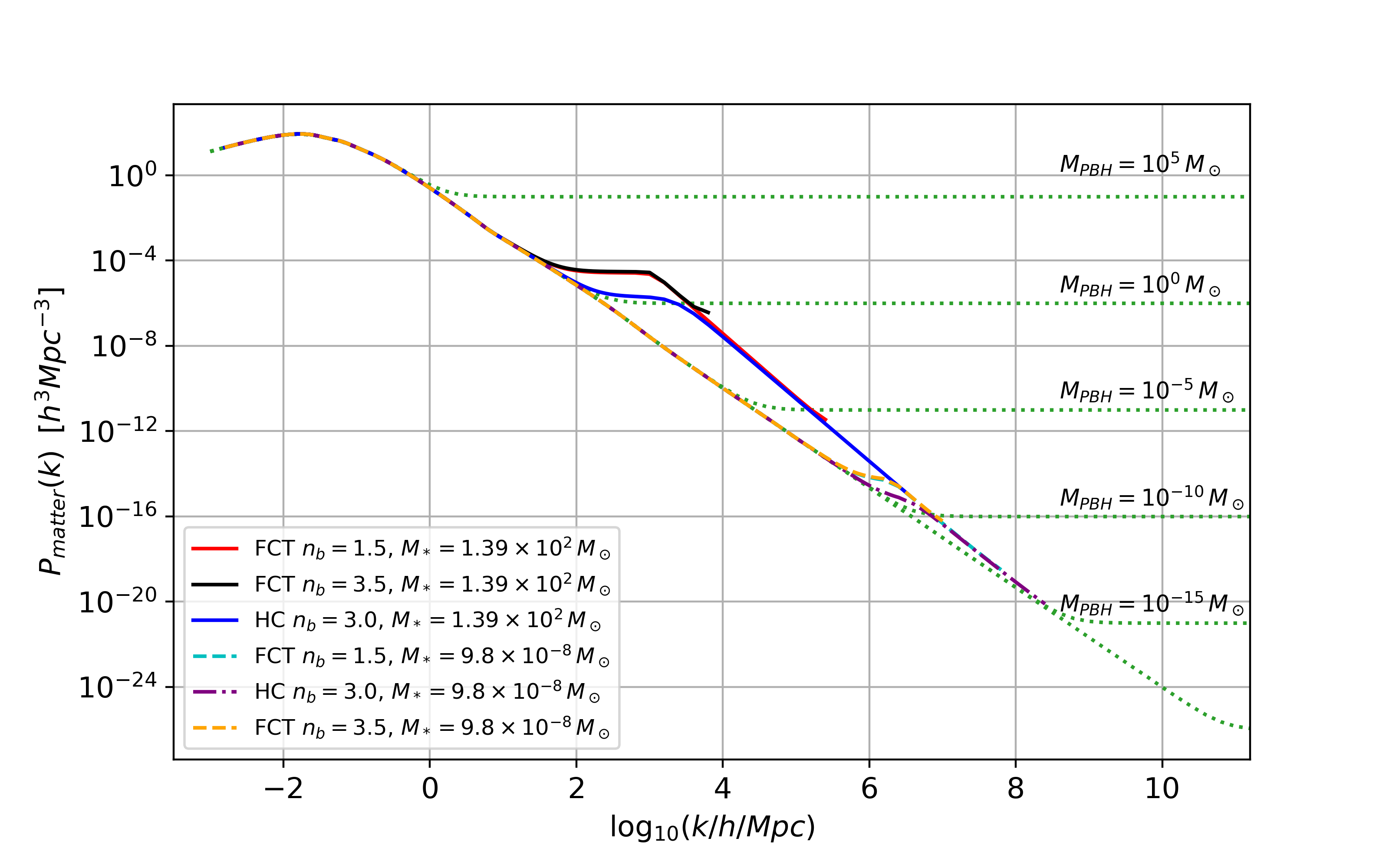}
    \caption{Matter power spectra at $z=20$ including noise in the total matter due to Poisson acting on the extended mass functions for the three scenarios considered in this work. Solid lines correspond to mass functions with characteristic mass $M_* = 1.39\times 10^{2}\, M_\odot$, while dashed/dot-dashed lines correspond to $M_* = 9.1\times 10^{-8}\, M_\odot$. Blue curves correspond to the horizon crossing (HC) scenario, with spectral blue index $n_b = 3.0$. Red curves correspond to the fix conformal time (FCT) scenario, with $n_b = 1.5$ and black curves correspond also to a FCT formation scenario, with $n_b = 3.5$. Green dotted lines correspond to the monochromatic power spectra with different PBH mass values. In this figure, we have assumed that $f_{\tiny{\mbox{PBH}}}=1$. Note that the knee in the power spectra at high wavenumbers corresponds to the Poisson noise from the discreteness of PBHs.}
    \label{Matter-Spectra}
\end{figure}

\subsection{The matter power spectrum}

As introduced before, the matter power spectrum is defined as the Fourier transform of the matter over-density field correlation function. This corresponds to the excess matter accumulation (or \textit{excess power}) at characteristic wave-numbers
$k = \frac{2\pi}{\lambda}$, which are related to modes with wavelength $\lambda$ in the
Fourier decomposition of $\delta(\vec
{x})$. The form of the power spectrum depends on physics at different
epochs. In particular, the primordial power spectrum is sourced by the physics
of \textit{inflation}, where most of the inflationary models predict a scale-invariant curvature power spectrum, i.e, it has a constant amplitude, regardless of scale. This causes the density power-spectrum to have a \textit{Harrison-Zeldovich} form, characterized by%
\begin{equation}
P_{\text{prim}}(k)  = A_{s}\,k^{n_{s}},
\end{equation}
where $A_{s}$ is the amplitude of the power-spectrum measured at any pivot wave-number $k_{0}$ and $n_s$ is the spectral index. Here we assume that $A_{s}=2.1\times 10^{-9} \rm Mpc^{3}$ (as measured by Planck at $k_{0}=0.05\,\text{Mpc}^{-1}$) and $n_{s}\approx 1$. Deviations from $n_{s}=1$ (perfect scale-invariance) are a
direct imprint of the slow-roll and acceleration parameters of inflationary
models \citep{Dizgah:2019, Arya:2020}. The measured power spectrum, however, is not the primordial one, as
during the cosmic evolution, a remarkable growth of structures at all scales occur due to gravitational interactions. This growth is parametrized in terms of the \textit{structure growth factor}, $D_{1}(z)$, which depends on the cosmic epoch during which the growth takes place (i.e. radiation dominance, matter
dominance, etc.). Likewise, the power of the different modes (characterized by the 
wave-number $k$) gets processed by \textit{sub-horizon physics}, which includes all
other effects such as gravitational interactions between over-dense regions as well as interactions with baryons, among others \citep{Ma:1995ApJ,PeacockBook} All these effects are parametrized in terms
of the \textit{transfer function} $T(k,z)$. Finally, as dark matter is (posited to be) made of discrete massive entities (i.e., primordial black holes), there is a further effect that has to be accounted for in the power spectrum. This last effect corresponds to the \textit{Poisson sampling noise}, which is currently being studied \citep{Padilla20}.

All in all, the matter power spectrum is given by
\begin{equation}
P_{\tiny{\text{matter}}}(k,z)  = A_{s}\,k^{n_{s}}\,T^{2}(k,z)D_{1}^{2}(z) + P_{\tiny{\text{Poisson}}}(k,z),
\end{equation}
where the different terms and coefficients (except for $P_{\tiny{\text{Poisson}}}(k,z)$) are given (for example) by the results of the Planck
collaboration, as well as from CMB Boltzmann codes like CAMB \citep{Lewis:2000}. Figure (\ref{Matter-Spectra}) shows the corresponding matter spectra for the particular cases considered in this work.

\section{Mechanisms for PBH generated primordial magnetic fields}
\label{sec:3}

In this section we present two possible mechanisms for generating primordial magnetic fields from PBH distributions. The first of them corresponds to the Biermann battery which, as pointed out in the introduction, has been proposed recently as a candidate for generating magnetic seed fields from primordial black hole distributions. Here we compute the field generated by a PBH of mass $M$ and maximal spin, which is supposed to be the largest possible within this mechanism, as well as argue why the total magnetic field generated by the whole PBH distribution should track the matter over-density field. Afterwards, we turn our attention to analyze the physics of magnetic PBHs, also recently proposed by Maldacena. With an analogous procedure, we start computing the contribution of the magnetic field generated by one PBH with a certain quantity of magnetic monopoles. We continue by discussing the relationship between the whole magnetic field perturbation coming from the total PBH distribution and the $\delta(\vec{x})$ field for this particular process.

\subsection{Biermann battery}

Magnetic field generation by the Biermann battery mechanism was actively studied in the past \citep{Hanayama05,Biermann,Bhattacharjee20} being nowadays one of the most prominent processes for justifying the small-scale seed cosmic magnetic field produced by thin accretion disks surrounding rotating black holes. One of the first studies of this mechanism applied in cosmological problems was first carried out for proto-circumstellar disks, based on radiation hydrodynamical simulations in two spatial dimensions \citep{Shiromoto14}. The main characteristics of this process can be summarized as follows: (i) for the battery to work, the electron density and electron pressure gradients cannot be co-linear (a situation which naturally holds for thin accretion disks); (ii) the  seed  magnetic  field grows linearly with time; (iii) the generation of the seed magnetic field depends only very weakly on the accretion rate (implying that no magneto-rotational instability (MRI) is necessary to operate, a priory, for the battery to work); (iv) there can be magnetic field generation in initially unmagnetized accretion discs around PBHs (which provide the small-scale seeds of cosmic magnetic field); (v) for thin accretion disks around rotating PBHs, the field generation rate increases when increasing the PBH spin, so the maximum seed field is expected to be generated for maximally rotating PBHs.

The magnetic field generated by the accretion disc surrounding the PBH corresponds (mainly) to a \textit{dipole} configuration, and therefore it has a fall-off $\propto r^{-3}$. At a fixed $r/r_{\tiny{\text{isco}}}$, being $r$ the radial distance from the PBH and $r_{\tiny{\text{isco}}}$ the radius of the innermost stable circular orbit of the PBH, the battery scales as $\sim M^{- 9/4}$, where $M$ is the mass of the PBH. In particular, we assume that for any given $M$ within the range of interest, an accretion disk around the PBH can be formed. Then, we can obtain the magnitude of the magnetic field for a PBH of mass $M$ and spin $s$ at a distance $r$ from its center as
\begin{equation}
B(M,s,r) = B_{\tiny{\text{isco}}}(M,s) \left(\frac{r_{\text{isco}}}{r}\right)^3,
\end{equation}
where $B_{\tiny{\text{isco}}}(M,s)$ is the value of the magnetic field generated by a black hole with mass $M$ and spin $s$, measured at $r_{\text{isco}}$. Also, as a first approximation, we assume that the BHs are \textit{near-extremal} \citep{Mirbabayi20}; i.e., $s=GM/c^2$, and therefore both $B_{\tiny{\text{isco}}}$ and $B$ no longer depend on the BHs spin parameter but depend on their masses. Within this description, the generated magnetic field is expected to provide an upper bound for this mechanism, which can then be compared with observations.

It is interesting to understand the reason why the seed magnetic field generated by the whole PBH distribution should track $\delta(\vec{x})$ for this particular process. The simplest assumption is that matter inhomogeneities become fluctuation sources for other fields, down to some \textit{correlation parameter}. One example where this would be appropriate for vector fields such as the magnetic field, $\vec{B}$, is the following. As it might seem intuitive, wherever there is a matter over-density, accumulation of PBHs is expected, each of which is surrounded by an accretion disk with some angular momentum $\vec{J}$ with random orientation. 
Also, it would likely be the case that there is some \textit{spatial correlation} between the alignments of those spins \citep{Ganeshaiah18}, as the over-densities source their own primordial torques over their own size-scales. Thus, wherever an accumulation of PBHs (and thus of dark matter) is present, a net magnetic field summed over \textit{coherently} should also exist. On the other hand, neighboring over-densities may lead to coherent addition of magnetic fields, but in the \textit{opposite} direction. Thus, choosing any particular component of the magnetic field; say, the one which is along the $z$ direction with respect to the plane of the disk), it is expected that \textit{half} of the over-densities will add to its net magnitude, whereas the subsequent half of them will substract from it. Therefore, if present, this phenomenon is automatically taken into account, in a \textit{purely statistical sense}, by considering that the whole magnetic field from the PBH distribution does track $\delta(\vec{x})$. Nonetheless, we do not discard other possibilities for the correlation between $\delta(\vec{x})$ and $B$. Here we provide only this simple example.

The contribution to the magnetic field generated by each PBH, taking as a reference value the one produced at a distance of $4\,r_{\tiny{\text{isco}}}$ from its center \citep{Biermann}, at redshift $z$, is
\begin{equation}\label{eq:Bierman-1PBH}
B_{\tiny{\text{Bier}}}(M,|\vec{x}^{\prime}-\vec{x}|, z) = \frac{\mathcal{C} B_{\tiny{\text{B}}}(M)}{a(z)^3}\left(  \frac{4\,r_{\tiny{\text{isco}}}(M)}{|\vec{x}^{\prime}-\vec{x}|}\right)^3,
\end{equation}
where $\mathcal{C}$ is a parameter that accounts for the correlation degree of the constituents, $a(z)$ is the cosmological scale factor at redshift z, the interval $|\vec{x}^{\prime}-\vec{x}|$ is measured in comoving coordinates, and
\begin{equation}
    B_{\tiny{\text{B}}}(M) \approx 10^{-2}\left[\frac{\scriptsize{\text{Gauss}}}{s}\right]\left(\frac{M}{5.0\, M_\odot}\right)^{-9/4}\left(\frac{G M}{\left(4\,r_{\tiny{\text{isco}}}\right)^3}\right)^{-1/2}.
\end{equation}

\subsection{Magnetic PBHs}

We now turn our analysis to study the plausibility of seed magnetic field generation coming from magnetic PBHs. A \textit{magnetic black hole} should be understood as a solution of the Standard Model coupled to gravity, and basically it describes a black hole with certain magnetic charge. Although magnetic charges do not require any new physics, they do not seem easy to produce, even less to observe in the Universe. One reason might be that magnetic monopoles are much heavier than electrically charged particles, implying that they are rather harder to pair create. Also, an electrically charged black hole can be neutralized within a conductive medium, which is not the case for magnetic black holes. Moreover, while it is possible to produce uncharged black holes \citep{Garcia-Bellido96, Garcia-Bellido17}, producing charged black holes seems harder.  During inflation, for instance, magnetic black hole production may seem unlikely \citep{Bousso96}. Nevertheless, it has been pointed out before that one possible mechanism is to first produce a large number of  monopoles  and  anti-monopoles, consider those primordial fluctuations that were able to collapse producing black holes and, if the black hole captures $N$ monopoles or anti-monopoles, then a net (statistical) magnetic charge of order $\sqrt{N}$ should be expected \citep{Bai20, Stojkovic05}. Moreover, if their masses are small enough, those PBHs would evaporate quickly becoming extremal black holes, that can survive until today, which were even proposed as plausible dark matter candidates \citep{Turner82,Bai20}.

As discussed in \cite{Maldacena20}, magnetic monopoles behave as point-like magnetic charges, giving rise to a net magnetic field analog to the Coulomb field, namely
\begin{equation}
\vec{B}_i(M,\,\vec{r}-\vec{r}_{i})  =\frac{\mu_{0}}{4\pi}\frac{q_{m}}{\left(\vec{r}-\vec{r}_{i}\right)  ^{3}}\left(\vec{r}-\vec{r}_{i}\right),
\end{equation}
where $q_{m}$ is the charge of a magnetic monopole, whose unit value is obtained from direct Dirac quantization \citep{Dirac31}, and in terms of fundamental constants it reads
\[
    q_m = 2\pi\,\frac{\epsilon_0 \hbar c^2}{e}.
\]
We consider the capture of monopoles to be instantaneous, such that a PBH of mass $M$ absorbs all the magnetic monopoles contained within its lagrangian volume, assuming an almost constant monopole density in the Universe. In this scenario, due to the fact that the presence of
monopoles with either positive or negative magnetic charge is equally
probable, the net magnetic charge of a PBH will follow a \textit{poissonian} distribution such that it will be well approximated by the square-root of the total number of absorbed monopoles, as pointed out before.

As for the Biermann battery mechanism, it is possible to justify why the total magnetic field perturbation from the whole PBH distribution should track $\delta(\vec{x})$ for this particular process. As the net magnetic charge follows a poissonian distribution (and thus each process can be though of as a 1-dimensional random walk with fixed step in  charge space), the net magnetic
charge $Q_{m}$ is approximately $q_{m}\sqrt{N}$, where $N$ is the total number of accreted monopoles. Thus, considering that magnetic monopoles are, roughly, \textit{homogeneously distributed} throughout the universe, and assuming that their number density $n_{\text{mon}}$ is known, it follows that the net magnetic charge of a region of space with volume $V$ is $\sqrt{n_{\text{mon}}\,V}$. Now, assuming that the magnetic monopoles \textit{trace} the PBH distribution, which is a reasonable assumption as they are more likely to be concentrated in the same regions where PBHs do form, one automatically gets that the net magnetic charge $Q_{m}$ of a region of (fixed) volume $V$ turns out to be \textit{proportional} to $\sqrt{\rho_{{}_{\tiny{\text{PBH}}}}}$, and thus
\begin{align}
Q_{m} & \propto  \sqrt{\rho_{{}_{\tiny{\text{PBH}}}}} \nonumber\\
& =\sqrt{\overline{\rho}_{\tiny{\text{DM}}}}\left(1+\delta(\vec{x})\right)^{1/2} \nonumber\\
& \approx \sqrt{\overline{\rho}_{\tiny{\text{DM}}}} \left(1 + \frac{1}{2}\delta(\vec{x})\right),
\end{align} 
for small $|\delta(\vec{x})|$. Finally, we recover a null average magnetic charge for the universe, as it is equally likely for a region to have either positive
or negative magnetic charge. Then, it is possible to \textit{displace} the magnetic charge distribution found before, such that it has zero average while preserving the
same fluctuation structure, to find that the distribution of net magnetic charge in the universe (always assuming the \textit{linear} regime, for which $\left\vert
\delta(\vec{x})\right\vert \ll 1$) is, indeed, \textit{proportional} to
$\delta(\vec{x})$. Thus, the total magnetic field perturbation coming from the whole PBH distribution should also track $\delta(\vec{x})$ for this particular magnetic monopole capturing process. 

From the previous discussion, we see that the magnitude of the magnetic field at distance $r$ away from a unique PBH of mass $M$ is found to be
\begin{equation}
B(M,r) = \frac{K\,\sqrt{M}}{r^{2}},
\end{equation}
where the orientation of the field is in the radial direction, but with equal probability of pointing either towards the PBH or away from it. The above equation implies that
\begin{equation}\label{eq:B-Maldacena}
B_{\tiny{\text{Mald}}}(M, |\vec{x}^{\prime}-\vec{x}|, z) = \frac{K\,\sqrt{M}}{a(z)^2\,|\vec{x}^{\prime}-\vec{x}|^2},
\end{equation}
where a(z) is the cosmological scale factor at redshift z and the interval $|\vec{x}^{\prime}-\vec{x}|$ is measured in comoving coordinates. 
Also, it is possible to deduce that the constant $K$ is given by
\begin{equation}\label{eq:K-nmonop}
K = \frac{\mu_{0}}{4\pi} q_{m}\, \sqrt{\frac{n_{\text{mon}}}%
{\rho_{\tiny{\mbox{PBH}}}}}\,,
\end{equation}
where $n_{\text{mon}}$ is the average number density of magnetic monopoles in
the universe and $\rho_{\tiny{\mbox{PBH}}}$ is the average cosmic mass density of
PBHs. 

\subsubsection{Magnetic monopole density}

We give estimates on the magnetic monopole density today from information at the PBH formation epoch, assuming they are all magnetic PBHs. To do so, we first use that the PBH density at the formation epoch is proportional to  the critical density of the universe at  PBH formation, $\rho_c$. Namely,
\begin{equation}
\rho_{\tiny{\mbox{PBH}}}(  a_{\text{form}})  =f_{\tiny{\mbox{PBH}}}\,f_{m}\,\rho_{\text{c}}\left(\left\langle a_{\text{form}}\right\rangle _{M\frac{dn}{dM}}\right),
\end{equation}
where $f_{\tiny{\mbox{PBH}}}$ is the fraction of dark matter in PBHs, $f_{m}$ is the fraction of energy in dark matter and $\left\langle a_{\text{form}}\right\rangle _{M\frac{dn}{dM}}$ is the average scale factor at the formation time. For said formation scale factors, we used values of $\,2\times10^{-26}$ and $4\times 10^{-12}$ in the FCT and HC scenarios, respectively, as discussed in \citet{Sureda:2020}. Thus, using equation (\ref{eq:K-nmonop}), the number density of magnetic monopoles at the time of PBH formation reads 
\begin{equation}\label{eq:nmonop-form}
n_{\text{mon}}\left(  a_{\text{form}}\right)  =\left(  \frac{4\pi}{\mu
_{0}q_{m}}\right)  ^{2}K^{2}f_{\tiny{\mbox{PBH}}}\,f_{m}\,\rho_{\text{c}}\left(\left\langle a_{\text{form}}\right\rangle _{M\frac{dn}{dM}}\right).
\end{equation}
The fraction $f_{m}$ is given by
\begin{equation}
f_{m}=\frac{\rho_{\tiny{\mbox{DM}}}\left(  \left\langle a_{\text{form}}\right\rangle_{M\frac{dn}{dM}}\right)}{\rho_{\text{c}}\left(\left\langle a_{\text{form}}\right\rangle_{M\frac{dn}{dM}}\right)}
\end{equation} 
and, taking into account that the critical density at PBH formation is well approximated by 
\begin{equation}
\rho_{\text{c}}\left(  \left\langle a_{\text{form}}\right\rangle _{M\frac
{dn}{dM}}\right)  \simeq\frac{\rho_{\text{rad,0}}}{\left(  \left\langle
a_{\text{form}}\right\rangle _{M\frac{dn}{dM}}\right)  ^{4}},
\end{equation} 
as well as that the dark matter density at PBH formation is given by
\begin{equation}
\rho_{\text{DM}}\left(  \left\langle a_{\text{form}}\right\rangle _{M\frac
{dn}{dM}}\right)  \simeq\frac{\rho_{\text{DM,0}}}{\left(  \left\langle
a_{\text{form}}\right\rangle _{M\frac{dn}{dM}}\right)^{3}},
\end{equation}
being $\rho_{\text{rad,0}}$ and $\rho_{\text{DM,0}}$ today's radiation and dark matter mass densities, respectively, the number density of monopoles \textit{at any} $a>a_{\text{form}}$ can be computed as
\begin{equation}\label{eq:nmonop-arbitrary-time}
n_{\text{mon}}\left(  a\right)  =n_{\text{mon}}\left(  a_{\text{form}}\right)
\left(\frac{a_{\text{form}}}{a}\right)^{3}.
\end{equation}
It is therefore interesting that, by fixing the value of the constant $K$
through comparison with the observed bounds in cosmic magnetic fields, it could be
possible to obtain bounds on the number density of magnetic monopoles today (which, of course, have not been observed yet). In fact, using the above expression for $n_{\text{mon}}$, we get $n_{\text{mon,0}}=n_{\text{mon}}\left(
a=1\right)$. Thus, considering that magnetic monopoles have a
mass $m_{\tiny{\mbox{GUT}}}$ \citep{Georgi74,Coleman73,Kiselev90}, we can directly translate the constraints on the
number density of magnetic monopoles today into constraints on $\Omega_{\text{mon,0}}$, i.e. the \textit{fractional contribution of monopoles to the current energy density budget}, by the relation
\begin{equation}\label{eq:nmonop-today}
\Omega_{\text{mon,0}}=\frac{m_{\tiny{\mbox{GUT}}}\,n_{\text{mon,0}}}{\rho_{\text{c,0}}%
},
\end{equation}
where $\rho_{\text{c,0}}$ is the critical density at $z=0$. Bearing in mind these two processes for magnetic field generation from primordial black holes, the next step is to study the statistics that arise when considering populations of PBHs with certain characteristics. In particular, we are interested in doing statistics from particular mass functions describing PBH populations at different epochs which could be applied for any other physical mechanism of magnetic field generation from primordial black hole distributions.

\section{Cosmic magnetic field generated from PBH distributions}
\label{sec:4}

We now propose a direct mechanism for understanding how cosmic magnetic field generation from a given PBH distribution at a certain epoch might be, based on physical field generation mechanisms within each PBH, and characterizing the further statistics over the whole distribution. 

As a first hypothesis, it is rather reasonable to assume that, since the Universe is homogeneous and isotropic at the largest scales, local orientations of the magnetic field are randomly aligned,
implying that the cosmic average, $\overline{B}$, of the magnetic field at any epoch is identically zero,
\begin{equation}
\overline{B}=0.
\end{equation}
Thus, any observed cosmic magnetic field should be a ``higher-order'' effect. Given that it is expected that the most prominent observed contribution to the total magnetic field should be the \textit{linear} contribution; i.e., the one coming from \textit{first-order} fluctuations in the matter distribution, we shall start by focusing on this one. 

We conjecture the following statement: \textit{Let $B_{i}(  M,|\vec{x}^{\prime}-\vec{x}|,z)$ be the magnitude of the contribution to the magnetic field at position $\vec{x}$ due to one PBH of mass $M$ located at position $\vec{x}^{\prime
}$, at redshift $z$. Then, the cosmic magnetic field contribution due to a (comoving) PBH distribution $\mbox{d}n/\mbox{d}M$ at a given point $\vec{x}$ of the cosmic volume and at redshift $z$ is given by the following integral formula:
\begin{multline}
\label{eq:master-formula}
\delta B_{i}(\vec{x},z)=\int_{0}^{\infty}dM\int_{\mathbb{R}^{3}}d^{3}{\vec{x}^{\prime}}\frac{dn}{dM}B_{i}\left(M,|\vec{x}^{\prime}-\vec
{x}|,z\right)\times\\  
\times S\left(\frac{\left\vert \vec{x}^{\prime}%
-\vec{x}\right\vert }{a(z)d(M)}\right)
\delta(\vec{x}^{\prime},z),
\end{multline}
where $S$ is some smooth cut-off function and $\delta(\vec{x},z)$ is the matter over-density field}.

The motivation of formula (\ref{eq:master-formula}) is the following. Considering that there are $\mbox{d}n/\mbox{d}M$ PBHs with masses between $M$ and $M+dM$, such that those located in the vicinity of $\vec{x}^{\prime}$ are at distance $\left\vert\vec{x}^{\prime}-\vec{x}\right\vert$ from the ``observation point''
$\vec{x}$, formula (\ref{eq:master-formula}) represents a sum over the whole PBH distribution of the magnetic field contribution due to each constituent. We further assume that, as expected, the magnetic field magnitude decays as
$\propto 1/r^{p}$ with respect to the distance $r$ to the observation point. In particular, the following form for the magnetic field generated by a source at position $\vec{x}$ due to a physical mechanism $i$ is considered:
\begin{equation}\label{eq:B-gral-form}
B_{i}(M,|\vec{x}^{\prime}-\vec{x}| , z) = \frac{C_{i}(M)}{\left\vert \vec{x}^{\prime}-\vec{x}\right\vert
^{p}},
\end{equation}
where the power $p$ depends on the dominant term in the associated magnetic field multipole expansion, and $C_i(M)$ is a function of $M$ which we shall particularize for each mechanism in later sections. For example, for the Biermann battery, expression (\ref{eq:Bierman-1PBH}) follows from (\ref{eq:B-gral-form}) by taking $C_i(M) = \mathcal{C} B_{\tiny{\text{B}}}(M)\left(4\,r_{\tiny{\text{isco}}}\right)^3$ and $p=3$. Similarly for the case of magnetic PBHs, we have that $C_{\tiny{\text{Mald}}}(M) = K\,\sqrt{M}/2$ (a factor $1/2$ arises from the fact that $\sqrt{\rho_{\tiny{\mbox{DM}}}}$ scales as $\delta(\vec{x})/2$, as explained in Section 3.2), and $p=2$ in the general expression for the magnetic fluctuation $\delta B_i$. The cut-off function $S$ is chosen as the following short-distance softening exponential function:
\begin{equation}
S\left(\frac{\left\vert\vec{x}^{\prime}-\vec{x}\right\vert}{d(M)}\right) = 1 - e^{-\left(  \frac{1.6\left\vert\vec{x}^{\prime}-\vec{x}\right\vert}{d(M)}\right)^3},
\end{equation}
which was obtained by Montecarlo simulations. Finally, we just integrated the resulting power spectrum over the wavenumber ranges associated to different magnetic field observations, to get the theoretical r.m.s. value corresponding to that particular scale. Note that the magnetic field perturbation is cut-off by the function $S$ at short distances. This is understood as a consequence of the fact that, typically, almost no PBH of mass $M$ is expected to be found within a comoving radius of $\frac{d(M)}{2}$ from the observation point. 

The most subtle point in formula (\ref{eq:master-formula}) concerns the factor $\delta(\vec{x}^{\prime}, z)$. We consider that $\delta B_i$ is proportional (at every point $\vec{x}^{\prime}$) to $\delta(\vec{x}^{\prime}, z)$, implying that magnetic field fluctuations linearly track matter density fluctuations due to reasons that depend on each process, as justified in the previous section.

Although the matter over-density field $\delta(\vec{x})$ takes values ranging from $-1$ to infinity, since purely linear perturbation are being assumed, it usually holds that $\left\vert \delta(\vec{x})\right\vert\ll 1$. 
Then, under-dense regions subtract from the average of $\delta$, whereas over-dense
regions increase it, in such a way that $\left\langle\delta(\vec{x})\right\rangle _{\vec{x}\in R^{3}} = 0$, by construction. Thus, over-dense regions positively contribute to the net $\vec{B}$ in the same direction to the direction of $\vec{B}$ at field point $\vec{x}$, whereas under-dense regions contribute with net $\vec{B}$ in the opposite direction. If the spatial distribution of under-dense regions has the same statistics as the matter spatial distribution of over-dense regions, then it is statistically valid to associate \textit{under-densities} with \textit{negative} $B$ and \textit{over-densities} with \textit{positive} $B$. This point makes it obvious that $\delta B_i$ \textit{is not} the value of the magnitude of $\vec{B}$ at point $\vec{x}$, but it is only a statistical tool to derive the clustering statistics and fluctuation spectra
of field perturbations, so it has a purely \textit{statistical} meaning, being able to give an estimate on its order of magnitude. We therefore shall refer to the field $\delta B_i$ as the \textit{statistical cosmic magnetic field}.

Finally, some $a(z)$ factors come up along the definitions, which are in place such that all distances are physical (proper) distances, instead of comoving ones. In what follows, we use the statistical magnetic field to derive properties of the primordial magnetic field fluctuations, as well as its corresponding power spectrum.

\subsection{The statistics of magnetic field fluctuations}

We start by rewriting the expression (\ref{eq:master-formula}) conjectured for the statistical magnetic field perturbation in terms of only comoving distances. As we shall see later on, this makes it easier to use the already known statistics of the matter density field introduced in Section \ref{sec:2}.

In the comoving frame, and including the form (\ref{eq:B-gral-form}) for the magnetic field generated by each PBH source, we get
\begin{multline}\label{eq:deltaB-comoving}
\delta B_i(\vec{x},z)=\int_{\mathbb{R}^{3}}d^{3}{\vec{x}^{\prime}}\left(
\int_{0}^{\infty}dM\frac{dn}{dM}\frac{C_{i}(M)  }{a(z)^{p}%
\left\vert \vec{x}^{\prime}-\vec{x}\right\vert ^{p}}\times\right.\\
\left.\times S\left(
\frac{\left\vert \vec{x}^{\prime}-\vec{x}\right\vert }{d(M)}\right)  \right) \delta(\vec{x}^{\prime}, z).
\end{multline}
Then, we notice that the quantity in parenthesis has compact support as $S_i$ has such a property, and it can be interpreted as a window-like function with a cut-off that goes as $\propto 1/r^p$, allowing to re-express (\ref{eq:deltaB-comoving}) as
\begin{equation}\label{eq:dB-F-delta}
\delta B_i (\vec{x}, z) = \int_{\mathbb{R}^{3}} F_{B}\left(\left\vert \vec{x}^{\prime}-\vec{x}\right\vert ,z\right)  \delta(
\vec{x}^{\prime}, z)  d^{3}\vec{x}^{\prime},
\end{equation}
where the kernel function $F_{B}$ is given by
\begin{equation}\label{eq:FB-def}
F_{B}(r,z)=\int_{0}^{\infty}\frac{dn}{dM}\frac{C_{i}(M)}{a(z)^{p}\,r^{p}}S\left(\frac{r}{d(M)}\right)\,dM,
\end{equation}

In analogy to the way we proceeded for the derivation of the matter power spectrum, we smoothly restrict the statistical $B$ field over a spatial scale of some size $R>0$, getting
\begin{equation}
\delta B_R^i(\vec{x}, z)=
\frac{1}{|\mathcal{B}_R|}\int_{\mathbb{R}^3}{W_R(\vec{x}^{\prime}-\vec{x})\,\delta B_i(\vec{x}^{\prime}, z)\,\mbox{d}^{3}\vec{x}^{\prime}},
\end{equation}
where $W_R(\vec{x}^{\prime}-\vec{x})$ is
the inverse Fourier transform of a spherical top-hat function in momentum space, multiplied by $|\mathcal{B}_R|$ (in order for it to be dimensionless). Once again, by the convolution formula we get
\begin{equation}
\widehat{\delta B}^i_R(\vec{k}, z) = \widehat{W}_R(\vec{k})\,\widehat{\delta B}_i(\vec{k}, z),
\end{equation}
where, as usual
\begin{equation}
\widehat{W}_R(\vec{k}) = \Theta\left(|\vec
{k}| - \frac{2\pi}{R}\right),
\end{equation}
which is dimensionless.
Then, from Parseval's theorem, we have for the variance that
\begin{align}
\sigma_{B}^{2}(R, z) & = \left\langle |\delta B_R^i(\vec{x}, z)|^{2}\right\rangle_{V}\nonumber \\
&  =\frac{1}{|V|}\int_{\mathbb{R}^3}
|\widehat{W}_{R}(\vec{k})|^{2}\;|
\widehat{\delta B}_i(\vec{k}, z)|^{2}\,\mbox{d}^{3}\vec{k}\nonumber\\
& = \int_{\mathbb{R}^3}
|\widehat{W}_R(\vec{k})|^{2}\;P_{B}(\vec{k}, z)\,\mbox{d}^{3}\vec{k}.
\end{align}
The quantity $\sigma_{B}(R, z)$ is interpreted as the root-mean-square value of the primordial magnetic field smoothed over scales of size $R$ at redshift $z$, whereas $P_{B}(\vec{k}, z)$ turns out to be, by definition, the \textit{statistical magnetic power spectrum}, which is then given by
\begin{equation}
P_{B}(\vec{k}, z) = \frac{1}{|V|}|
\widehat{\delta B}_i(\vec{k}, z)|^{2}.
\end{equation}
and $\widehat{\delta B}_i(\vec{k}, z)$ is the Fourier transform of the statistical magnetic field perturbation given by expression (\ref{eq:master-formula}). In fact, by means of equation (\ref{eq:dB-F-delta}) and by the convolution formula, this can be expressed as the product between the Fourier transform of $F_B(r,z)$ and $\delta(\vec{k},z)$, namely
\begin{equation}
\widehat{\delta B}_i(\vec{k}, z) = \widehat{F_{B}}(\vec{k}, z)\;  \widehat{\delta}(\vec{k}, z).
\end{equation}
Finally, the magnetic power spectrum is
\begin{align}
P_{B}(\vec{k}, z) &= \frac{1}{|V|}|\widehat{\delta B}_i(
\vec{k},z)|^{2}\nonumber \\
& = \frac{1}{|V|}|\widehat{F_{B}}(\vec{k}, z)|^{2}\;|\widehat{\delta}(\vec{k}, z)|^{2} \nonumber\\
& = |\widehat{F_{B}}(\vec{k}, z)|^{2}\;P_{\Lambda\scriptsize{\mbox{CDM}}}(\vec{k}, z),
\end{align}
where $P_{\Lambda\scriptsize{\mbox{CDM}}}(\vec{k}, z)$ is the $\Lambda$CDM-matter power spectrum.

\subsection{The square-averaged magnetic field and the cosmic magnetic energy
density}

The magnetic energy density must be proportional to the square of the magnetic field at any point in the cosmic volume. In a statistical sense, the magnitude of the average cosmic magnetic field should be understood as the root mean square of the magnetic field fluctuations, namely
\begin{equation}
    B \simeq \sqrt{\left\langle |\delta B_i|^2\right\rangle},
\end{equation}
where the square-averaged magnetic field, which is proportional to the cosmic magnetic energy density in the universe, is given by
\begin{equation}
\left\langle |\delta B_i|^2\right\rangle =\frac{1}{|V|}\int_{\mathbb{R}^3}{
|\delta B_i(\vec{x}, z)|^{2}\,\mbox{d}^{3}\vec{x}}.
\end{equation}
Then, using Parseval's theorem, we have that%
\begin{align}
\left\langle |\delta B_i|^{2}\right\rangle & =\frac{1}{|V|} \int_{\mathbb{R}^3}{|\widehat{\delta B}_i(\vec{k}, z)|^{2}\;
\mbox{d}^{3}\vec{k}} \nonumber \\
& = \int_{\mathbb{R}^3}{
P_{B}(\vec{k}, z)\;\mbox{d}^{3}\vec{k}}.
\end{align}
Therefore, the cosmic magnetic energy density at redshift $z$ is given by
\begin{align}
\rho_{B}(z) & = \frac{\left\langle |\delta B_i|^{2}\right\rangle}{2\mu_{0}}\nonumber \\
& = \frac{1}{2\mu_{0}}\int_{\mathbb{R}^3}{P_{B}(\vec{k}, z)\;\mbox{d}^{3}\vec{k}},
\end{align}
where $\mu_{0}$ is the vacuum magnetic permeability. Thus, all the interesting statistical properties of $\delta B_i$ are characterized by means of its power spectrum $P_{B}$, which in turn depends on $\widehat{F_{B}}$ (the Fourier transform of the window-like function) and the matter power spectrum $P_{\Lambda\scriptsize{\mbox{CDM}}}$ which is already known. 

We now give a closed explicit expression for $\widehat{F_{B}}$, which shall be computed afterwards for the particular cases of Biermann's battery and Maldacena's magnetic black holes. We start by taking expression (\ref{eq:FB-def}) and rewrite it in the following way:
\begin{equation}
F_{B}(|\vec{x}^{\prime}-\vec{x}|',\, z) = \int_{0}^{\infty}{
\frac{dn}{dM}\frac{C_{i}(M)}{a(z)^p}\,G_i(|\vec{x}^{\prime}-\vec{x}|,\,M)\,\mbox{d}M},
\end{equation}
where the function $G_i(r)$ is given by
\begin{equation}
    G_i(r,M) = \frac{S(r/d(M))}{r^p}
\end{equation}
and captures all the spatial dependence of $F_B$.
Then, by linearity of the Fourier transformation, we have that
\begin{equation}
\widehat{F_{B}}(\vec{k}, z) = \int_{0}^{\infty}{\frac{dn}{dM} \frac{C_i(M)}{a(z)^p}\widehat{G}_i(\vec{k},M)\,\mbox{d}M},
\end{equation}
where
\begin{equation}
\widehat{G}_i(\vec{k},M) = \int_{\mathbb{R}^3}{
\frac{S\left(\left\vert \vec{x}\right\vert/d(M)\right)  }{\left\vert \vec{x}\right\vert ^p}e^{-i\vec{k}\cdot\vec{x}}\;\mbox{d}^{3}\vec{x}}.
\end{equation}
As shown in Appendix B, we find that
\begin{equation}
\widehat{G}_i(\vec{k},M) = \frac{4\pi}{k}\int_0^{\infty}{\frac{S\left(r/d(M)\right)  }{r^{p-1}}\sin(kr)\,\mbox{d}r}.
\end{equation}
Therefore, we finally get
\begin{multline}
\widehat{F_{B}}(\vec{k}, z)  = \frac{4\pi}{k}\int_0^{\infty}{\mbox{d}M}\int_0^{\infty}\mbox{d}r\;\frac{dn}{dM}\,\frac{C_i(M)}{a(z)^p}\times\\
\times\frac{S(r/d(M))}{r^{p-1}}\sin(|\vec{k}|r).
\end{multline}

\begin{figure*}
    \centering
    \includegraphics[scale=0.6]{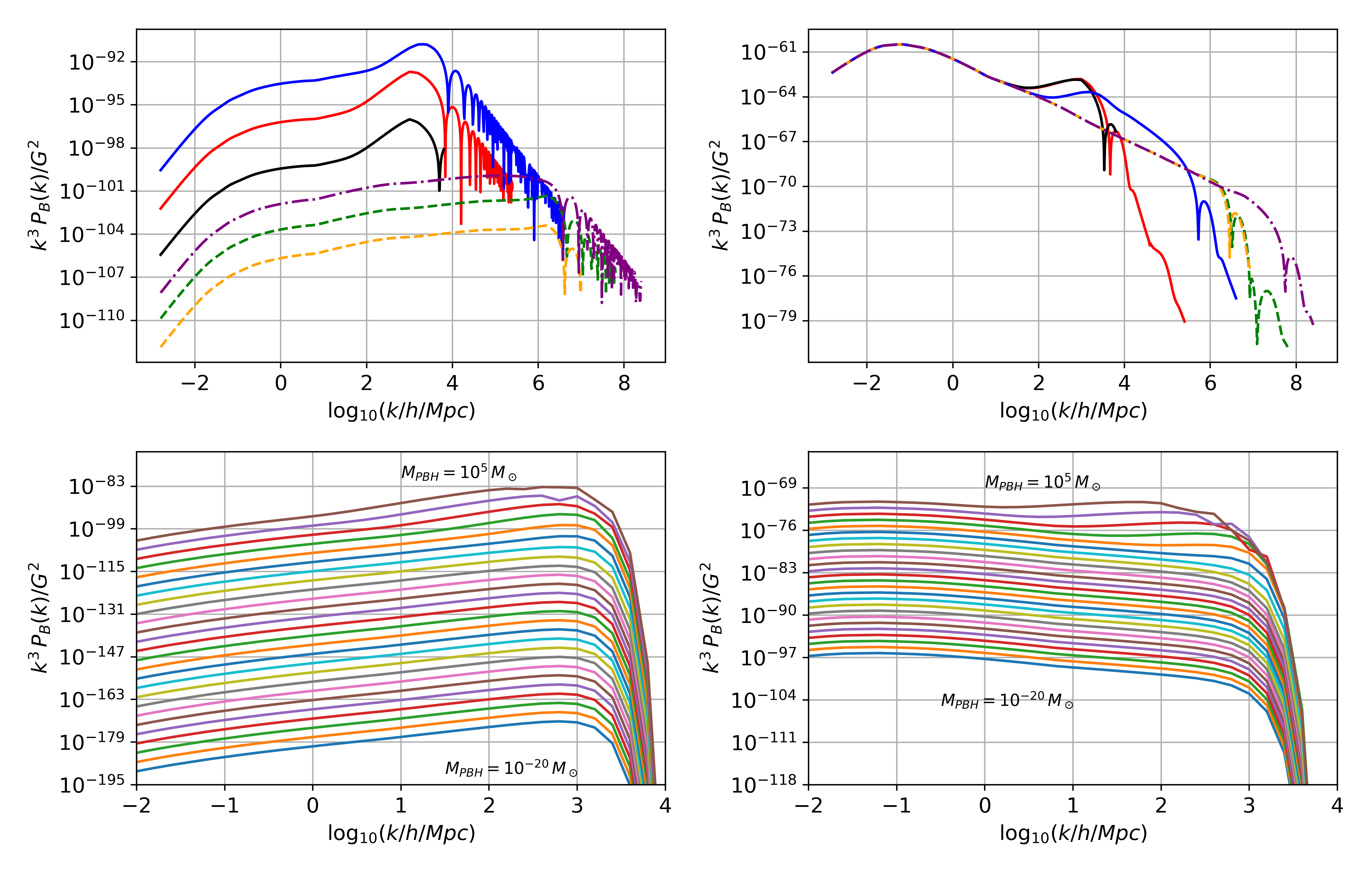}
    \caption{Statistical seed magnetic field power spectrum generated by PBH distributions at redshift $z = 20$. \textit{Top left panel:} Magnetic field power spectra assuming each PBH with a Biermann battery turned on, for different scenarios. Solid lines correspond to characteristic mass $M_* = 1.39 \times 10^{2}\,M_\odot$. The blue curve corresponds to the HC formation scenario, with spectral index $n_b = 3.0$, while the red and black ones correspond to the FCT formation scenario, with spectral indices of $n_b = 1.5$ and $n_b = 3.0$ respectively. Dotted lines correspond to PBH mass functions with characteristic mass $M_* = 9.1 \times 10^{-8}\,M_\odot$. The purple curve corresponds to the HC formation scenario (with $n_b=3.0$) and the green and orange curves correspond to the FCT scenario (with $n_b = 1.5$ and $n_b = 3.5$ respectively). \textit{Top right panel:} Power spectra of a r.m.s. averaged seed magnetic field of about $10^{-30}$ G at $z=20$, produced by magnetic PBHs. Solid lines (with the same convention of colors as in the case of the Biermann battery) correspond to PBH mass functions with characteristic mass $M_* = 1.39 \times 10^{2}\,M_\odot$, while dotted curves correspond to the case with $M_* = 9.1 \times 10^{-8}\, M_\odot$. \textit{Bottom panels:} PBH mass sweep comparison from $10^{-20}\, M_\odot$ to $10^{5}\, M_\odot$ for seed magnetic field power spectra for the Biermann battery (left panel) and magnetic PBHs (right panel) considering monochromatic mass functions.}
    \label{fig:magnetic-spectra}
\end{figure*}

This general expression can be evaluated numerically for a given functional form of the mass function $\frac{dn}{dM}$, and for each magnetic field generation mechanism.

\begin{table*}
\centering
\begin{tabular}{|c|c|c|c|c|c|}
\cline{4-6}
\multicolumn{3}{c|}{} & \centering \textit{Biermann battery}  & \multicolumn{2}{c|}{\centering \textit{Magnetic PBHs \tiny{$\left(\sqrt{\langle|\delta B_i|^2\rangle}\sim 10^{-30}\, G \right)$}}}\\ 
\hline
$M_*\,[M_\odot]$ & \textit{Scenario} & $n_b$ & $\sqrt{\langle|\delta B_i|^2\rangle}\, [G]$  & $K\, \left[G\,\mbox{Mpc}^2\, M_\odot^{-1/2}\right]$ & $\Omega_{\tiny{\mbox{mon,0}}}$\\
\hline\hline
\multirow{3}{2cm}{\centering $1.39\times 10^2$} &
\multirow{2}{2cm}{\centering FCT} & $1.5$ & $2\times 10^{-47}$ & $3\times 10^{-44}$ & $2\times 10^{-16}$\\ %\cline{3-6}
&  & $\;3.5$ & $4\times 10^{-49}$ & $6\times 10^{-43}$ & $6\times 10^{-14}$\\ %\cline{2-6}
& HC & $3.0$ & $1\times 10^{-46}$ & $3\times 10^{-45}$  & $2\times 10^{-18}$\\
\hline
\multirow{3}{2cm}{\centering $9.1\times 10^{-8}$} &
\multirow{2}{2cm}{\centering FCT} & $1.5$ & $1\times 10^{-51}$ & $4\times 10^{-49}$ & $3 \times 10^{-26}$\\ %\cline{3-6}
&  & $3.5$ & $1\times 10^{-52}$ & $2\times 10^{-48}$ & $5 \times 10^{-25}$\\ %\cline{2-6}
& HC & $3.0$ & $8\times 10^{-51}$ & $1\times 10^{-49}$ & $3 \times 10^{-27}$\\ 
\hline
\end{tabular}
\caption{Summary of the numerical results for the different magnetic field generation scenarios considered in this work. \textit{Biermann battery:} root mean square of the seed magnetic field generated by this mechanism for different PBH mass functions. \textit{Magnetic PBHs:} values of the Coulombian constant in the formula for the magnetic field generated by accreted monopoles on each PBH, assuming a seed field of $10^{-30}$ G at redshift $z = 20$. In the last column, the $\Omega$ values for today's magnetic monopole energy density parameter.}
\label{table:results}
\end{table*}

Finally, if a monochromatic mass function is assumed, we get
\begin{equation}\label{eq:monochromatic-mass-function}
    \frac{dn}{dM} = \frac{f_{\tiny{\mbox{PBH}}}\,\rho_{\tiny{\mbox{DM}}}}{M_{\tiny{\mbox{PBH}}}}\,\delta\left(M - M_{\tiny{\mbox{PBH}}}\right),
\end{equation}
and, as noticed in Section 2, the characteristic distance $d(M)$ in the case of monochromatic functions gets reduced to $d(M) = n^{-1/3}$,
where 
\begin{equation}\label{eq:nc-monocrh}
    n = \frac{f_{\tiny{\mbox{PBH}}}\,\rho_{\tiny{\mbox{DM}}}}{M_{\tiny{\mbox{PBH}}}}
\end{equation}
Then, plugging this together with equations (\ref{eq:monochromatic-mass-function}) and (\ref{eq:nc-monocrh}) into the general expression for the magnetic fluctuation obtained in equation (\ref{eq:deltaB-comoving}), we get
\begin{equation}\label{eq:monochr-deltaBi}
    \delta B_i(\vec{x},z) = \int_{\mathbb{R}^3}{\mbox{d}^3\vec{x}^{\prime}\;\frac{C_i\left(M_{\tiny{\mbox{PBH}}}\right)}{a(z)^p\,|\vec{x}-\vec{x}^{\prime}|^p}\;S\left(n^{1/3}|\vec{x}-\vec{x}^{\prime}|\right)\delta(\vec{x}^{\prime},z)}.
\end{equation}
Also, the kernel function $F_B(\vec{x},z)$ defined in expression (\ref{eq:FB-def}) becomes
\begin{equation}
    F_B(\vec{x}-\vec{x}^{\prime},z) = \frac{C_i\left(M_{\tiny{\mbox{PBH}}}\right)}{a(z)^p}\,\left(\frac{S\left(n^{1/3}|\vec{x}-\vec{x}^{\prime}|\right)}{|\vec{x}-\vec{x}^{\prime}|^p}\right),
\end{equation}
and its Fourier transform reads
\begin{equation}
    \widehat{F}_B(\vec{k}
    ,z) = \frac{C_i\left(M_{\tiny{\mbox{PBH}}}\right)}{a(z)^p} \,\widehat{G}(\vec{k}, M_{\tiny{\mbox{PBH}}}),
\end{equation}
where, by means of the calculation shown in Appendix B,
\begin{equation}
    \widehat{G}(\vec{k}, M_{\tiny{\mbox{PBH}}}) = \frac{4\pi}{|\vec{k}|}\int_{0}^{\infty}{\mbox{d}r\,\frac{S\left(n^{1/3}\,r\right)}{r^{p-1}}\,\sin(|\vec{k}\,r|)}.
\end{equation}
All the following steps towards the magnetic power spectrum and the average magnetic energy are completely analogous to the case of continuous (extended) mass functions.

\section{Magnetic field and power spectra for different generation mechanisms}
\label{sec:5}

We now show numerical results from the formalism introduced in the previous section, applied to the Biermann battery and to magnetic PBHs.

We assume a universe at redshift $z = 20$ populated by PBHs of different masses, starting with Press-Schechter mass distribution functions within both fixed conformal time (FCT) and horizon crossing (HC) scenarios, introduced in Section 2. The simulations were performed considering two different characteristic masses, $M_* = 1.39 \times 10^{2} M_\odot$ and $M_* = 9.1\times 10^{-8} M_\odot$, and spectral blue indices $n_b = 1.5$ and $n_b = 3.5$ for FCT, and $n_b = 3.0$ for HC. This set of parameters was suggested in \citet{Sureda:2020}, for which all dark matter can be composed of PBHs (that is, $f_{\text{PBH}} = 1$). We assumed that PBHs that may have formed at a given redshift $z$ could have masses varying between the evaporated mass at that redshift, $M_{\tiny{\mbox{ev}}}(z)$, and the mass at which the cumulative PBH number density equals only one PBH per horizon volume, which we refer as $M_{1\tiny{\mbox{pH}}}$. This maximum mass turns out to be also a function of $z$, as the Hubble volume increases with cosmic time, being thus $M_{1\tiny{\mbox{pH}}}$ the maximum mass of PBHs which are in \textit{causal contact} at redshift $z$. 

The previous PBH mass range yields also an associated range in the frequency domain. In fact, taking into account the monotonic characteristic distance $d(M)$ introduced in Section 2, we considered the frequency as to range from $k_{min} = 2\pi/d(M_{1\tiny{\mbox{pH}}})$ to $k_{max}=2\pi/d(M_{\tiny{\mbox{ev}}})$. With this, we numerically computed the magnetic power spectrum, considering also the short distance cut-off function $S$ obtained from Montecarlo simulations. The unique constraint imposed in the case of the Biermann battery at redshift $z = 20$ was the maximality of the spin parameter of all PBHs.  Thus, our results report on the maximum expected statistical average magnetic field that could be produced by a PBH population, each one with a Biermann battery on. 

For the case of magnetic field generation from magnetic PBHs, we imposed observational constraints in order to set the seed field to the expected value at redshift $z=20$. With this, it was possible to estimate the magnetic monopole number density at such $z$, for then to extrapolate it to the corresponding monopole number density today, as computed in the previous section. The program is the following: once the observational constraint on the seed fields was set at $z = 20$ to be approximately $10^{-30}$ G, as reported in \citet{Ashoorioon05}, we use the corresponding magnetic field spectrum to estimate the coupling constant $K$ between the magnetic field produced by a magnetic PBH and its mass. Then, following equations (\ref{eq:nmonop-form}), (\ref{eq:nmonop-arbitrary-time}) and (\ref{eq:nmonop-today}), we estimate the magnetic monopole number density today, in terms of today’s radiation and dark matter mass densities and the critical density at PBH formation. Afterwards, the density parameter $\Omega_{\tiny{\mbox{mon,0}}}$ associated to such monopole number density was computed from equation (\ref{eq:nmonop-today}), and then compared with current upper limits on said density parameter.

The magnetic power spectra for both field generation mechanisms is shown in Figure \ref{fig:magnetic-spectra}, where the spectrum from the top left panel corresponds to the Biermann battery mechanism, and the one from the right panel corresponds to the case of magnetic PBHs. Each curve corresponds to a different formation scenario, all at $z=20$. Below them is shown a comparison of the corresponding magnetic field spectra if monochromatic mass functions were considered, scanning the characteristic mass from $10^{- 20} M_\odot$ to $10^{5} M_\odot$.

Finally, following the Planck collaboration \citep{Planck16}, which defines the magnetic field strength smoothed on a certain scale $\lambda$, namely
\begin{equation}
    \left\langle B^2_{\lambda}\right\rangle  = \frac{1}{2\pi^2}\int_0^{\infty}{P_B(k)\,k^2\,e^{-\lambda^2\,k^2}\,\mbox{d}k}
\end{equation}
we computed the average statistical magnetic seed at $z = 20$ at a scale $\lambda\sim 1\,\mbox{kpc}$, and the values were found to be the ones shown in Table \ref{table:results}. 

\begin{figure}
    \centering
    \includegraphics[scale=0.45]{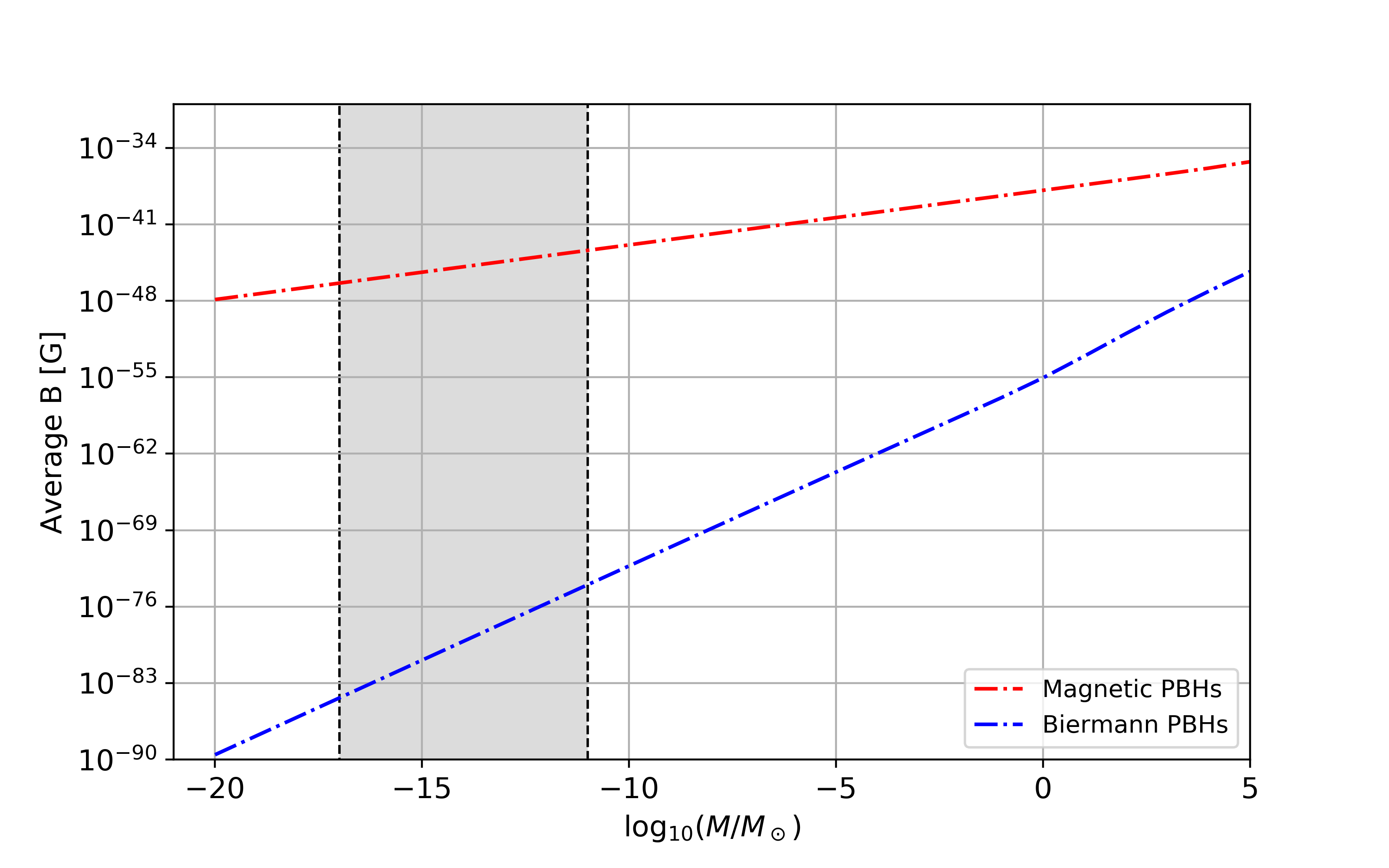}
    \caption{Maximum averaged seed magnetic field expected from PBH monochromatic mass functions. The blue curve corresponds to the Biermann battery generation mechanism and the red one to magnetic PBHs, assuming a maximum monopole density parameter of $\Omega^{\tiny{\mbox{max}}}_{\tiny{\mbox{mon,0}}} \sim 10^{-12}$, expected from observations \citep{Medvedev:2017JCAP}. The grey area indicates the PBH mass range for which all dark matter could have been formed from monochromatic distributions of PBHs.}
    \label{Monochromatic-MaxB}
\end{figure}

Finally, for a comparative analysis, we analyze the dependence of the maximum magnetic field that is expected to be generated for monochromatic mass functions, for each one of the mechanisms discussed in this work. For magnetic PBHs, we assumed the maximal magnetic monopole abundance at $z=20$ compatible with previous estimations, which was found to be $\Omega_{\tiny{\mbox{max}}} = 10^{-12}$ \citep{Medvedev:2017JCAP}. The results are shown in the Figure \ref{Monochromatic-MaxB}, from which a fairly linear behavior is observed. In the figure, the gray area indicates the PBH mass range for which all dark matter could have been formed from monochromatic distributions of PBHs. We note that for both generation mechanisms, the r.m.s. fields are less than what is required to account for measured magnetic fields after amplification, being $\approx$ 15 orders of magnitude lower for the case of Biermann battery and $\approx$ 5 orders of magnitude lower for magnetic PBHs.

\section{Discussion}
\label{sec:6}

We developed a general method for obtaining the fluctuation statistics and characteristic power spectra of PBH-generated magnetic fields at cosmological scales. We assumed that each PBH is capable of generating its own magnetic field by some physical mechanism, and that the characteristic fluctuations in the net magnetic field produced by the whole PBH population correlates linearly with the corresponding fluctuations in the dark matter density distribution, on the same scale. This is a robust formalism, which is independent on the physical generation mechanism. Assuming particular scenarios of PBH distributions given by monochromatic and Press-Schechter PBH mass functions, we explored two different magnetic field generation mechanisms. The first one is due to the Biermann battery mechanism, and the second one is due to the cosmic abundance of magnetic monopoles in the early universe, that could have been accreted by the PBH population, becoming magnetic PBHs.

We found that the Biermann battery mechanism, by itself, is not capable of producing the required seed field at $z=20$ in order to account, after evolution, for the present day magnetic fields in voids. That is, it is not possible that the whole PBH population generates a sufficiently large net magnetic field due to only a Biermann battery working on each PBH constituent. This estimation holds in the simplest case, that is, despite all PBHs, independently of their mass, having an accretion disk around them, being able to drive the corresponding Biermann induction mechanism. Furthermore, it is also assumed that the PBH spin is maximal for each PBH constituent, and that there is a perfect correlation between the magnetic field fluctuations and the density fluctuation field. By hydrodynamic considerations, it can be argued that it is unlikely for PBHs of masses lower than $10^{-22}\,M_{\sun}$ to form the required accretion disk, as the expected disk will be too faint, and thus not dense enough to be approximated by a fluid. 
Nonetheless, our estimates constitute an upper bound of the expected net seed field provided by this mechanism, and even considering the simplest case, it suffices to rule out the Biermann battery mechanism as a magnetic field generation mechanism to explain cosmic magnetic fields, in the sense addressed in this work.

Regarding the magnetic PBH mechanism, assuming that the required seed field at $z=20$ has been reached, we were able to deduce a density parameter $\Omega_{\tiny{\mbox{mon,0}}}$ for the magnetic monopole number density at the present day, which is safely within observational constraints, as reported in \cite{Medvedev:2017JCAP}. These estimates result in favor of the conjecture that primordial magnetic fields could be generated from magnetic PBHs, being consistent with the current measurements of cosmic magnetic fields.

As pointed out in Appendix A, assuming typical dynamo mechanisms for magnetic field amplification and regeneration in order to reach the present-day values measured in voids, we estimated that a seed field of around $10^{-30}\, G$, at $z = 20$, would be needed for reaching the present-day average magnetic field of order $10^{-16}$ G, quoted by \citet{Vachaspati:2020}. This estimation is in concordance with previous results, in particular the ones reported in \citet{Ashoorioon05}.

We find it relevant to point out that, when considering magnetic PBHs as the origin of cosmic magnetic fields, it should be studied whether or not dark matter properties or baryons in the Universe are affected in some way that might be inconsistent with observations. Namely, considering all the PBHs to be magnetically charged is equivalent to considering that the dark matter candidate is a population of magnetically charged, extremely massive particles, which no longer interact with baryonic matter through the gravitational interaction only, but also electromagnetically, by virtue of their magnetic field. Although the typical magnetic fields generated over cosmological scales by this process are small enough that they do not modify usual physical processes, the fields in the vicinity of these magnetically charged dark matter particles may be strong enough to change the way they interact with normal baryonic matter.

Finally, our general method could be implemented for assessing any number of tentative magnetic field generation mechanisms that could explain how a PBH generates a magnetic field in its vicinity, modeled by a particular functional form which depends on the PBH mass. We expect to use this method to explore other different magnetic field generation channels in the future.

\section*{Acknowledgments}

The work of I.J.A. is funded by ANID, REC Convocatoria Nacional Subvenci\'on a Instalaci\'on en la Academia Convocatoria A\~no 2020, Folio PAI77200097. M.E.R and F.A.S thank support by grants PIP 11220130100365CO and PICT-2016-4174, from CONICET and FONCyT (Argentina); and by SECyT-UNC. M.E.R is a postdoctoral fellow of CONICET (Argentina).
NP acknowledges support from Fondecyt 1191813. NP, JM and JS acknowledge support from CONICYT project Basal AFB-170002.

\section*{Data Availability}
No new data were generated or analysed in support of this
research.

\section*{Appendix A: A little detour on dynamo action in Void Galaxies}
\label{AppendixA}

To constrain the effects of the seeding in possible observations, we discuss how the different seeding models will compare with \citet{Neronov10}.
To do so, we follow \citet{Beck2013} to calculate the magnetic field filling factor in voids. Also, we rely on semi-analytic approximations  \citep{Rodrigues2019,Rodrigues2015} in order to estimate the evolution of the magnetic field from seeds to galaxies. 

It is possible to infer the dynamo action in typical galaxies in voids, where an e-folding time of $\tau_{MF} \sim 20$ Myr is expected. This allows for an increase on the typical (rms) magnitude of the cosmic magnetic field of 20 orders of magnitude from $z=20$. In the case of the {\it Biermann battery}, it is not sufficient to reach the corresponding equipartition values, and therefore to populate the intergalactic medium in voids.

However, for magnetic PBHs with extended PBH mass functions, it is possible to reach a magnetic field of about $0.001\, \mu G$ and therefore, the action of cosmic rays may populate the intergalactic medium to the observed values of the magnetic field. 

\section*{Appendix B: Supplementary calculations}
\label{AppendixB}

In this appendix, we calculate the Fourier transform of the ``shape function'' of the potential $G_M(\vec{x})$ used in section \ref{sec:4}, in order to obtain closed expressions for the magnetic power spectrum used in this work.

Let $G(\vec{x})$ be given by
\begin{equation}
G(\vec{x})  = \frac{S_{i}(\left\vert \vec
{x}\right\vert/d(M))}{\left\vert \vec{x}\right\vert^{p}}.
\end{equation}
Then, its 3D Fourier transform is given by
\begin{align}
\widehat{G}_{M}(\vec{k}) & = \int_{\mathbb{R}^3}
\mbox{d}^{3}\vec{x}\,\frac{S_{i}(\left\vert \vec
{x}\right\vert/d(M))}{\left\vert \vec{x}\right\vert^{p}}\,e^{-i\vec{k}\cdot
\vec{x}}\nonumber\\
&  = 2\pi \int_0^{\infty}\int_0^{\pi}
\mbox{d}r\,\mbox{d}\theta\,r^{2}\sin\theta\frac{S_{i}(r/d(M))}{r^{p}}\,e^{-ikr\cos\theta}\nonumber\\
&  =-2\pi\int_0^{\infty}\int_0^{\pi}
\mbox{d}r\,\mbox{d}(\cos(\theta))\,  r^{2}\frac{S_{i}(r/d(M))}{r^{p}}e^{-ikr\cos\theta}\nonumber\\
&  = 2\pi\int_0^{\infty}
\mbox{d}r\frac{S_{i}(r/d(M))}{r^{p-2}}\int_{-1}^{1}
\mbox{d}\xi\; e^{-ikr\xi}\nonumber\\
&  =\frac{2\pi}{ik}%
\int_0^{\infty}
\mbox{d}r\frac{S_{i}(r/d(M))}{r^{p-1}}\left(
e^{ikr}-e^{-ikr}\right)  \nonumber\\
&  =\frac{4\pi}{k}%
\int_0^{\infty}
\mbox{d}r\frac{S_{i}(r/d(M))}{r^{p-1}}%
\sin\left(  kr\right)  .
\end{align}
Thus, we finally get
\begin{equation}
\widehat{G}_{M}(\vec{k})  =\frac{4\pi}{|\vec{k}|}
\int_0^{\infty}
\mbox{d}r\frac{S_{i}(r/d(M))}{r^{p-1}}%
\sin\left(|\vec{k}|\,r\right)  .
\end{equation}

%%%%%%%%%%%%%%%%%%%%%%%%%%%%%%%%%%%%%%%%%%%%%%%%%%

%%%%%%%%%%%%%%%%%%%% REFERENCES %%%%%%%%%%%%%%%%%%

% The best way to enter references is to use BibTeX:

\bibliographystyle{mnras}
\nocite{*}
\bibliography{biblio}
{}

%%%%%%%%%%%%%%%%%%%%%%%%%%%%%%%%%%%%%%%%%%%%%%%%%%

%%%%%%%%%%%%%%%%% APPENDICES %%%%%%%%%%%%%%%%%%%%%

%%%%%%%%%%%%%%%%%%%%%%%%%%%%%%%%%%%%%%%%%%%%%%%%%%
%%%c
%%%% https://arxiv.org/pdf/astro-ph/0010373.pdf
%%% c

% Don't change these lines
\bsp	% typesetting comment
\label{lastpage}
\end{document}